\documentclass[aps,prx,twocolumn,notitlepage,10pt,longbibliography]{revtex4-2}

\usepackage{tikz}
\usepackage{tikz-cd}

\usepackage{graphicx}
\usepackage{amsfonts,amsmath,amssymb}
\usepackage{comment}
\usepackage[colorlinks=true]{hyperref}

\begin{document}

\title{Spontaneous symmetry breaking in the Heisenberg antiferromagnet on a triangular lattice}

\date{\today}

\author{Basti\'an Pradenas}
\author{Grigor Adamyan}
\author{Oleg Tchernyshyov}
\affiliation{
William H. Miller III Department of Physics and Astronomy, 
Johns Hopkins University,
Baltimore, Maryland 21218, USA
}
\date{\today}

\begin{abstract}
We present a detailed investigation of an overlooked symmetry structure in non-collinear antiferromagnets that gives rise to an emergent quantum number for magnons. Focusing on the triangular-lattice Heisenberg antiferromagnet, we show that its spin order parameter transforms under an enlarged symmetry group, $\mathrm{SO(3)_L \times SO(2)_R}$, rather than the conventional spin-rotation group $\mathrm{SO(3)}$. Although this larger symmetry is spontaneously broken by the ground state, a residual subgroup survives, leading to conserved Noether charges that, upon quantization, endow magnons with an additional quantum number---\emph{isospin}---beyond their energy and momentum. Our results provide a comprehensive framework for understanding symmetry, degeneracy, and quantum numbers in non-collinear magnetic systems, and bridge an unexpected connection between the paradigms of symmetry breaking in non-collinear antiferromagnets and chiral symmetry breaking in particle physics.
\end{abstract}

\maketitle

\section{Introduction}

In quantum mechanics and quantum field theory, quantum numbers serve as discrete labels assigned to eigenstates of observables, characterizing intrinsic properties, underlying symmetries, and associated conservation laws. While quantum numbers typically classify distinct states, degeneracies can arise—situations where multiple states share identical eigenvalues, such as energy. In these cases, additional quantum numbers or auxiliary labels are introduced to fully distinguish among degenerate states. One famous example is the \emph{isospin}, introduced by Heisenberg~\cite{heisenberg1989bau} to account for the nearly identical strong interactions experienced by neutrons and protons. By regarding the neutron and the proton as two isospin states of the same nucleon, one assigns them the isospin projections $I_3 = -\frac{\hbar}{2}$ and $I_3 = +\frac{\hbar}{2}$, respectively. This idea extends to all hadrons, including pions, which form an isospin triplet ($I=\hbar$). Pions posed a puzzle due to their anomalously light masses compared to other hadrons. The resolution came in the 1960s with the idea of \textit{spontaneous chiral symmetry breaking} in QCD, where the underlying chiral symmetry, $ \mathrm{SU(2)_L \times SU(2)_R} $, is broken down to its diagonal subgroup---identified as isospin rotations---thereby rendering pions as the (pseudo) Goldstone bosons in the limit of vanishing up and down quark masses. The low-energy physics of these Goldstone bosons is captured by \emph{non-linear sigma model} (NLSM), in which the isospin group emerges as the residual symmetry after chiral symmetry breaking~\cite{nambu1961dynamical, gell1960axial, lee1968phenomenological}.

In condensed-matter systems, particularly in ordered magnets, NLSMs serve as low-energy effective theories for describing symmetry breaking and magnetic excitations. For instance, in two-sublattice  antiferromagnets, the spontaneous breaking of the global spin-rotation symmetry, $\mathrm{SO}(3)$, down to $\mathrm{SO}(2)$ gives rise to two magnons as Goldstone bosons. The unbroken $\mathrm{SO}(2)$, which corresponds to spin rotations about the ordered spins, serves as a residual symmetry that allows one to define the magnon \emph{spin projection} quantum number, often denoted as $S_{z}$ assuming the magnetic order aligns with $\mathbf{e}_{z}$. Various authors have referred to this quantum number as ``pseudo-spin,'' ``polarization,'' or ``helicity''~\cite{proskurin2017spin, daniels2018nonabelian, qaiumzadeh2018controlling, yu2018polarization, kamra2020antiferromagnetic, wimmer2020observation, guckelhorn2023observation}. 

A more intriguing case emerges in non-collinear antiferromagnets, such as those on triangular lattices. There, the non-collinear spin order breaks the global spin $\mathrm{SO}(3)$ symmetry, leaving \emph{no obvious axis} for continuous rotations and suggesting no remaining residual symmetry. Yet one finds that two of the three magnon branches are degenerate \cite{dombre1989nonlinear, chernyshev2009spin, dasgupta2020theory, pradenas2024spin}, hinting at an underlying residual symmetry. 

In this paper, we resolve this apparent paradox by reexamining the symmetry structure of the spin order parameter of non-collinear antiferromagnets---represented by an $\mathrm{SO(3)}$ matrix $O$---in fact supports a larger set of transformations than is commonly emphasized. Specifically, $O$ transforms under two independent rotation groups, $\mathrm{SO(3)_L}$ and $\mathrm{SO(2)_R}$, acting from the left and right, respectively, as previously noted in \cite{PhysRevLett.64.3175,PhysRevLett.70.2483}. The right-acting group $\mathrm{SO(2)_R}$ corresponds to rotations that leave the spin plane orientation invariant. As a result, the full symmetry group is $\mathrm{SO(3)_L \times SO(2)_R}$, rather than a single $\mathrm{SO(3)}$.

A detailed analysis of the spontaneous breaking of this group by the non-collinear spin order reveals that a residual symmetry subgroup survives. This residual symmetry (1) preserves the non-collinear ground state, (2) underlies the degeneracy observed in two of the magnon branches, and (3) endows the magnons with an additional quantum number---\emph{isospin}. We adopt this term from chiral symmetry breaking in QCD, where a similar residual symmetry gives rise to the isospin quantum number for pions. 

The paper is structured as follows. In Sec.~\ref{sec:SSBNeel}, we briefly revisit the collinear Néel antiferromagnet in its $\mathrm{O}(3)$-NLSM formulation. We highlight how its spontaneous symmetry breaking yields magnons with a well-defined spin projection. In Sec.~\ref{sec:NoncollinearNLSM}, we move to the triangular-lattice Heisenberg antiferromagnet, where the order parameter is naturally described by $\mathrm{O}(4)$-NLSM. We then clarify why two spin-wave branches remain degenerate despite the complete breaking of $\mathrm{SO}(3)$. In Sec.~\ref{sec:Chiral}, we introduce the enlarged symmetry group—the chiral group $\mathrm{SO(3)_L \times SO(3)_R}$—and describe its action on the spin order parameter, establishing it as the full symmetry group relevant to the problem. In Sec.~\ref{sec:O(4)-formulation}, we present the $\mathrm{O}(4)$-NLSM formulation of the triangular antiferromagnet. This parametrization proves especially convenient, as the chiral group transformations act directly on vectors rather than on matrices. In Sec.~\ref{sec:Chiral Symmetry Breaking}, we detail how the ground state of the non-collinear spin order spontaneously breaks the chiral symmetry and identify the subgroups that remain unbroken. In Sec.~\ref{sec:isospin}, we use these residual symmetries to construct the associated Noether charges and introduce the isospin quantum number—an additional quantum label for magnons, beyond their energy and momentum. Finally, in Sec.~\ref{sec:Discussion}, we summarize our results and comment on possible extensions to other non-collinear magnets.


\section{NLSM for the Néel AFM}
\label{sec:SSBNeel}

In a two-sublattice antiferromagnet (AFM), the order parameter is the N\'eel vector $\mathbf{n}$, defined as the difference in magnetization between the two sublattices,
\begin{equation}
\mathbf{n} = \frac{1}{2}\left(\mathbf{m}_{1} - \mathbf{m}_{2}\right),
\end{equation}
where $\mathbf{m}_1$ and $\mathbf{m}_2$ are unit vectors representing the directions of the magnetization of sublattices 1 and 2, respectively. Each sublattice spin is given by
\begin{equation}
\mathbf{S}_{a} = S \, \mathbf{m}_{a}, \qquad a=1,2,
\end{equation}
with $S$ denoting the spin length. Additionally, a uniform magnetization field is defined by
\begin{equation}
\mathbf{m} 
= \mathbf{m}_1 + \mathbf{m}_2.
\end{equation}
Although $\mathbf{m}$ vanishes in a ground state, it becomes nonzero during antiferromagnetic dynamics. At low-energies, $\mathbf{m}$ is a slave of the Néel vector and is often written as
\begin{equation} \label{magnetization-dynamics-Neel}
\mathbf{m} =\chi\mathcal{S}\,\mathbf{n}\times\partial_{t}{\mathbf{n}},
\end{equation}
where $\chi$ is the magnetic susceptibility, and $\mathcal{S} = S/A_{\mathrm{uc}}$ is the spin density, defined by the total spin $S$ per magnetic unit cell of area $A_{\mathrm{uc}}$.

\begin{figure}[ht]
    \centering
    \includegraphics[width=0.40\textwidth]{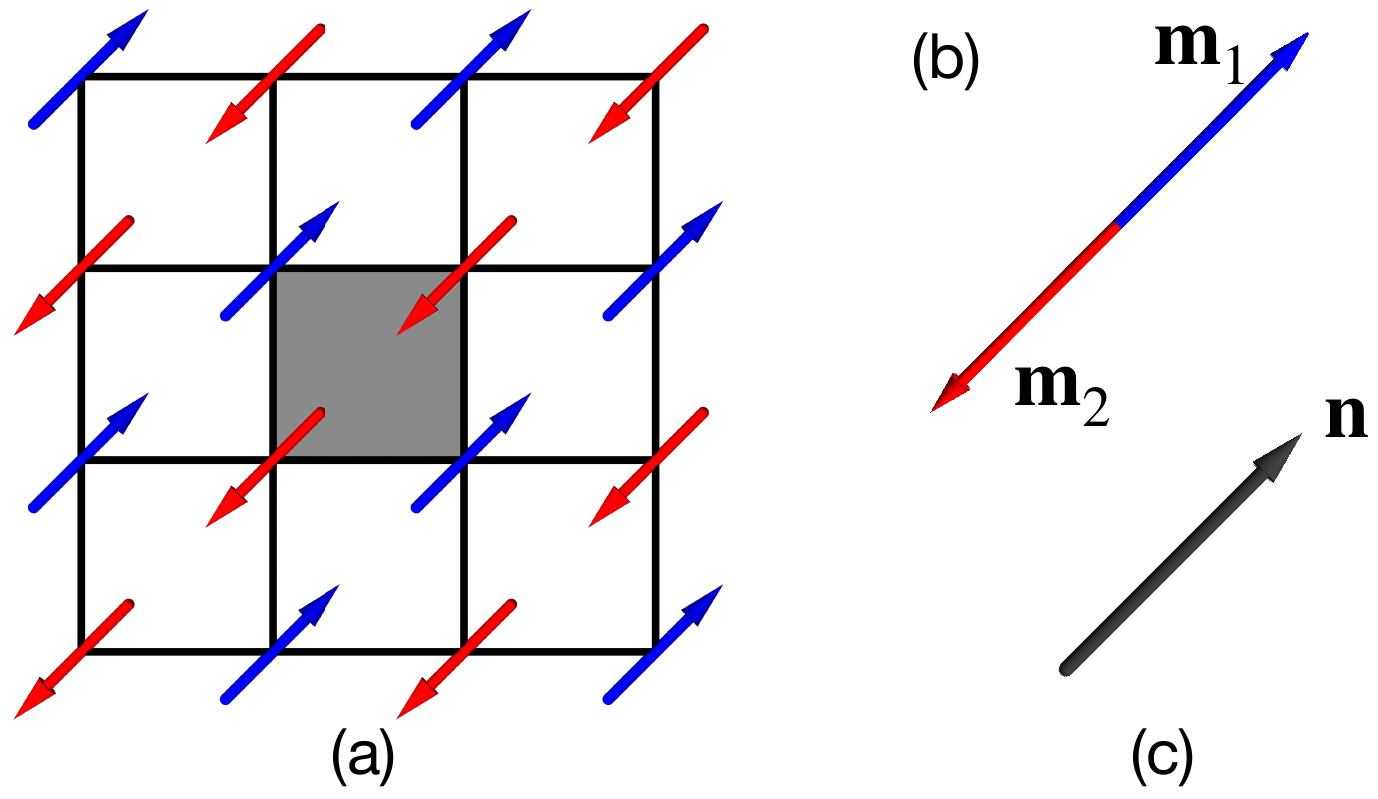}
    \caption{(a) Illustration of the antiferromagnetic (Néel) order, where the shaded region indicates the magnetic unit cell. (b) Schematic representation of the magnetic order parameter. (c) The corresponding Néel order vector.}
    \label{fig:neel}
\end{figure}

The continuum field theory of the Heisenberg Néel antiferromagnet can be succinctly formulated as an $\mathrm{O}(3)$ nonlinear sigma model (NLSM)~\cite{Baryakhtar:1979a, Andreev:1980, Haldane:1983}, with the Lagrangian given by:
\begin{equation} \label{eq:Neel-NLSM}
\mathcal{L}
= \frac{\rho}{2}
\partial_t \mathbf{n} \cdot \partial_t \mathbf{n}
-
\frac{\mathcal{J}}{2}\,
\partial_i \mathbf{n}\cdot \partial_i \mathbf{n},
\end{equation}
where $\rho = \chi \mathcal{S}^2$ is the inertia density, $\mathcal{J}$ is the exchange stiffness~\cite{dasgupta2020theory, pradenas2024spin}, and summation over repeated indices is assumed. The latin indices $i$ span over the dimensions $x, y$. This Lagrangian is invariant under the global spin-rotation, the $\mathrm{SO}(3)$ group, resulting in the Noether current
\begin{equation}
\label{neel-noether}
\mathbf{j}^{t}
=\rho
\mathbf{n} \times \partial_{t}{\mathbf{n}},
\quad
\mathbf{j}^{i} =-\mathcal{J}
\mathbf{n} \times \partial_{i}\mathbf{n}.
\end{equation}
Its conservation directly encodes the equations of motion for this system. The time component $\mathbf{j}^t$ is the spin density, related to the uniform magnetization,
\begin{equation} \label{Neel-spin-density}
\mathbf{j}^{t}=\mathcal{S}\mathbf{m}.
\end{equation}

\subsection{Spontaneous Symmetry Breaking}

The ground state of the Néel antiferromagnet spontaneously breaks the global $\mathrm{SO}(3)$ symmetry down to $\mathrm{SO}(2)$, allowing the Néel vector to point in an arbitrary direction in three-dimensional space. The continuous family of symmetry-related ground states thus corresponds to points on the unit sphere $S^2$. We denote the ground-state orientation by $\mathbf{n}_0$, and without loss of generality, choose it to point along the $z$-axis: $\mathbf{n}_0 = \mathbf{e}_z$. To analyze this symmetry breaking, we introduce the generators of the Lie algebra of the rotation group $\mathrm{SO}(3)$, denoted by $S_i$, with $i = x, y, z$. The generators $S_x$ and $S_y$ correspond to rotations about axes perpendicular to the ground state and are therefore spontaneously broken, while $S_z$, corresponding to rotations about the ground state direction $\mathbf{e}_z$, remains unbroken.

Magnons, the Goldstone bosons arising from this spontaneous symmetry breaking, correspond to small transverse deviations of the spin order parameter from the ground state orientation $\mathbf{n}_0$. These small deviations can be parametrized as

\begin{equation}
\mathbf{n} = \left( n_x, n_y, \sqrt{1 - n_x^2 - n_y^2} \right),
\end{equation}
where the transverse components $n_x$ and $n_y$ are small $n_x^2 + n_y^2 \ll 1$.

Since the vector $\mathbf{n}$ is constrained to lie on $S^2$, only two of its three components are independent. Consequently, the broken symmetry generators $S_x$ and $S_y$ induce nonlinear shift transformations on the Goldstone modes to preserve the unit-length constraint $|\mathbf{n}|=1$:
\begin{equation} 
\label{RxRy}
n_{a} \rightarrow n_{a} + \alpha_{a} \sqrt{1 - n_x^2 - n_y^2},
\end{equation}
where the Latin indices $a = x,y$ denote the transverse components, and $\alpha_{a}$ are infinitesimal angles associated with shifts in the directions orthogonal to the ground state direction $\mathbf{n}_{0}$.

In contrast, the Goldstone modes transform linearly under the unbroken $\mathrm{SO}(2)$ subgroup, corresponding to rotations about the $\mathbf{n}_{0}$ direction generated by $S_z$:
\begin{equation}
\label{Neel-SO(2)}
n_{a} \rightarrow n_{a} + \beta \epsilon_{ab} n_{b},
\end{equation}
here $\beta$ is an infinitesimal rotation angle and $\epsilon_{ab}$ is the antisymmetric Levi-Civita symbol in two dimensions.

\subsection{Linearized Theory and Noether Currents}

To calculate the Noether currents associated with these transformations, we expand the Lagrangian about the ground state $\mathbf{n}_0$ and introduce small amplitude spin waves via 
\begin{align} \label{eq:phi-doublet}
\delta \mathbf{n} = \boldsymbol{\phi} \times \mathbf{n}_0,
\end{align}
where the vector field doublet $\boldsymbol{\phi} = (\phi_x, \phi_y, 0)$ encodes the local rotation angles about directions transverse to the ground state. Substituting this into Eq.~\eqref{eq:Neel-NLSM}, we obtain:
\begin{equation}
\label{eq:Lin-Neel-NLSM}
\begin{aligned}
\mathcal{L} = \frac{\rho}{2} \Big( \partial_{t}\phi_x^2 &+ \partial_{t}\phi_y^2  \Big) - \frac{\mathcal{J}}{2} \Big( \nabla \phi_x^2 + \nabla \phi_y^2  \Big),
\end{aligned}
\end{equation}
where the gradient operator is defined as $\nabla = (\partial_x, \partial_y)^T$. At linear order, the transformation in Eq.~\eqref{RxRy} simplifies to a linear shift, $\phi_{a} \rightarrow \phi_{a} + \alpha_{a}$. The quadratic Lagrangian, Eq.~(\ref{eq:Lin-Neel-NLSM}), remains invariant under these field shifts, which directly lead to the conservation of Noether currents:
\begin{align}
   j^{t}_{a} = \rho\, \partial_{t} \phi_{a}, \quad j^{i}_{a} = -\mathcal{J}\, \partial_{i}\phi_{a}.
\end{align}
These results are nothing but the linear-order Noether currents of Eq.~(\ref{neel-noether}). 
Furthermore, the Lagrangian in Eq.~(\ref{eq:Lin-Neel-NLSM}) preserves the symmetry under global SO(2) rotations (Eq.~\eqref{Neel-SO(2)}) of the fields $\boldsymbol{\phi} = (\phi_x, \phi_y, 0)$ about the vacuum state, reflecting the degeneracy of the two magnon branches with speed $c_{\mathrm{I,II}}=\sqrt{\rho/\mathcal{J}}$. This residual $\mathrm{SO}(2)$ symmetry ensures another conserved current:
\begin{align}\label{Neel-SO(2)-current}
    j^{t}_{\mathrm{SO(2)}} = \rho \epsilon_{ab} \phi_{a}\partial_{t}\phi_{b},\quad  j^{i}_{\mathrm{SO(2)}} = -\mathcal{J} \epsilon_{ab} \phi_{a}\partial_{i}\phi_{b},
\end{align}
where $a,b \in \{x,y\}$ and $\epsilon_{ab}$ is the two-dimensional Levi-Civita symbol. Physically, this current represents the spin density defined in Eq.~(\ref{Neel-spin-density}), projected along the ground-state direction $\mathbf{n}_0=\mathbf{e}_{z}$.

\subsection{Spin Projection as a Quantum Number}

Consider circularly polarized spin waves around the ground-state order $\mathbf{n}_0$. To analyze these modes systematically, we first introduce the complex fields
\begin{equation}
\phi_{\pm} = \frac{1}{\sqrt{2}}\Bigl(\phi_{x} \pm i\,\phi_{y}\Bigr),
\end{equation}
which compactly represent the small-angle doublet defined in Eq.~\eqref{eq:phi-doublet}A plane-wave solution for these fields can then be naturally expressed as
\begin{align}
\phi_{\pm}(x) = \frac{\phi_{0}}{\sqrt{2}}\, e^{\pm i(\omega t - \mathbf{k} \cdot \mathbf{x})},
\end{align}
where we have introduced the spacetime coordinate $x = (t, \mathbf{x})$, with $t$ denoting time and $\mathbf{x} = (x, y)$ the two-dimensional spatial position. Here, $\omega$ is the angular frequency, and $\mathbf{k} = (k_x, k_y)$ is the corresponding wavevector. The $+$ sign corresponds to a right-handed circular polarization, while the $-$ sign corresponds to a left-handed one. Each of these modes carries a definite spin-density projection along the ground-state axis $\mathbf{n}_0$
\begin{align}
    j^{t}_{\mathrm{SO(2)}} = \pm \rho \omega \phi_0^2.
\end{align}
Upon canonical quantization, we promote the fields $\phi_{\pm}(x)$ to operators and introduce the bosonic creation and annihilation operators $a_{\mathbf{k}}^{\dagger},a_{\mathbf{k}}$ and $b_{\mathbf{k}}^{\dagger},b_{\mathbf{k}}$ corresponding to the quanta of $\phi_{+}$ and $\phi_{-}$, respectively. These operators satisfy the bosonic commutation relations 
\begin{equation} [a_{\mathbf{k}},a_{\mathbf{k}'}^\dagger] = \delta_{\mathbf{k},\mathbf{k}'}, \quad [b_{\mathbf{k}},b_{\mathbf{k}'}^\dagger] = \delta_{\mathbf{k},\mathbf{k}'},
\end{equation} 
with all other commutators vanishing.
The operators expansions of the fields take the form
\begin{equation}
\label{eq:mode-expansion-cm}
\begin{split}
\phi_{+}(x)
&=
\sqrt{\frac{\hbar}{\rho A}}
\sum_{\mathbf{k}}
\frac{1}{\sqrt{2\omega}}
\Bigl(
b_{\mathbf{k}}\,
e^{-i k x}
+
a_{\mathbf{k}}^\dagger\,
e^{i k x}
\Bigr),
\\[4pt]
\phi_{-}(x)
&=
\sqrt{\frac{\hbar}{\rho A}}
\sum_{\mathbf{k}}
\frac{1}{\sqrt{2\omega}}
\Bigl(
a_{\mathbf{k}}\,
e^{-i k x}
+
b_{\mathbf{k}}^\dagger\,
e^{i k x}
\Bigr),
\end{split}
\end{equation}
where $A$ is the sample area, $kx = \omega t - \mathbf{k} \cdot \mathbf{x}$ and $\omega = \sqrt{\rho/\mathcal{J}}|\mathbf{k}|$. 

The total spin projection $S_z$ is obtained by integrating the conserved $\mathrm{SO}(2)$ current over space. Substituting the mode expansions of $\phi_{\pm}$ into $j^t_{\mathrm{SO(2)}}$ in Eq.~\eqref{Neel-SO(2)-current}, we obtain:
\begin{align} \label{eq:Sz-detailed}
S_{z}= \;\mathopen{:}
\int_{A} d^2x \; j^t_{\mathrm{SO(2)}}(x)\mathopen{:}
\;=
\hbar\,\sum_{\mathbf{k}} 
\left(
 a_{\mathbf{k}}^\dagger a_{\mathbf{k}}
-\,
 b_{\mathbf{k}}^\dagger b_{\mathbf{k}}
\right).
\end{align}
Here, the symbol $\mathopen{:}\mathclose{:}$ denotes normal ordering of operators. Defining the number operators for the quanta of $\phi_{+}$ and $\phi_{-}$ as
\begin{equation}
N_{+} = \sum_{\mathbf{k}} a_{\mathbf{k}}^\dagger a_{\mathbf{k}}, \quad
N_{-} = \sum_{\mathbf{k}} b_{\mathbf{k}}^\dagger b_{\mathbf{k}},
\end{equation}
we have
\begin{equation}
S_{z} = \hbar\,\bigl(N_{+}-N_{-}\bigr).
\end{equation}
Thus, $\phi_+$ magnon quanta carry spin projection $S_z = \hbar$, while $\phi_-$ magnon quanta carry spin projection $S_z = -\hbar$. The residual $\mathrm{SO}(2)$ symmetry guarantees that each magnon is labeled not only by its energy and momentum, but also by a well-defined spin projection.

This example illustrates the intimate connection between residual symmetries and the conserved charges and quantum numbers—a connection that we will further exploit in the following sections.

\begin{figure}[ht]
\centering
\includegraphics[width=0.45\textwidth]{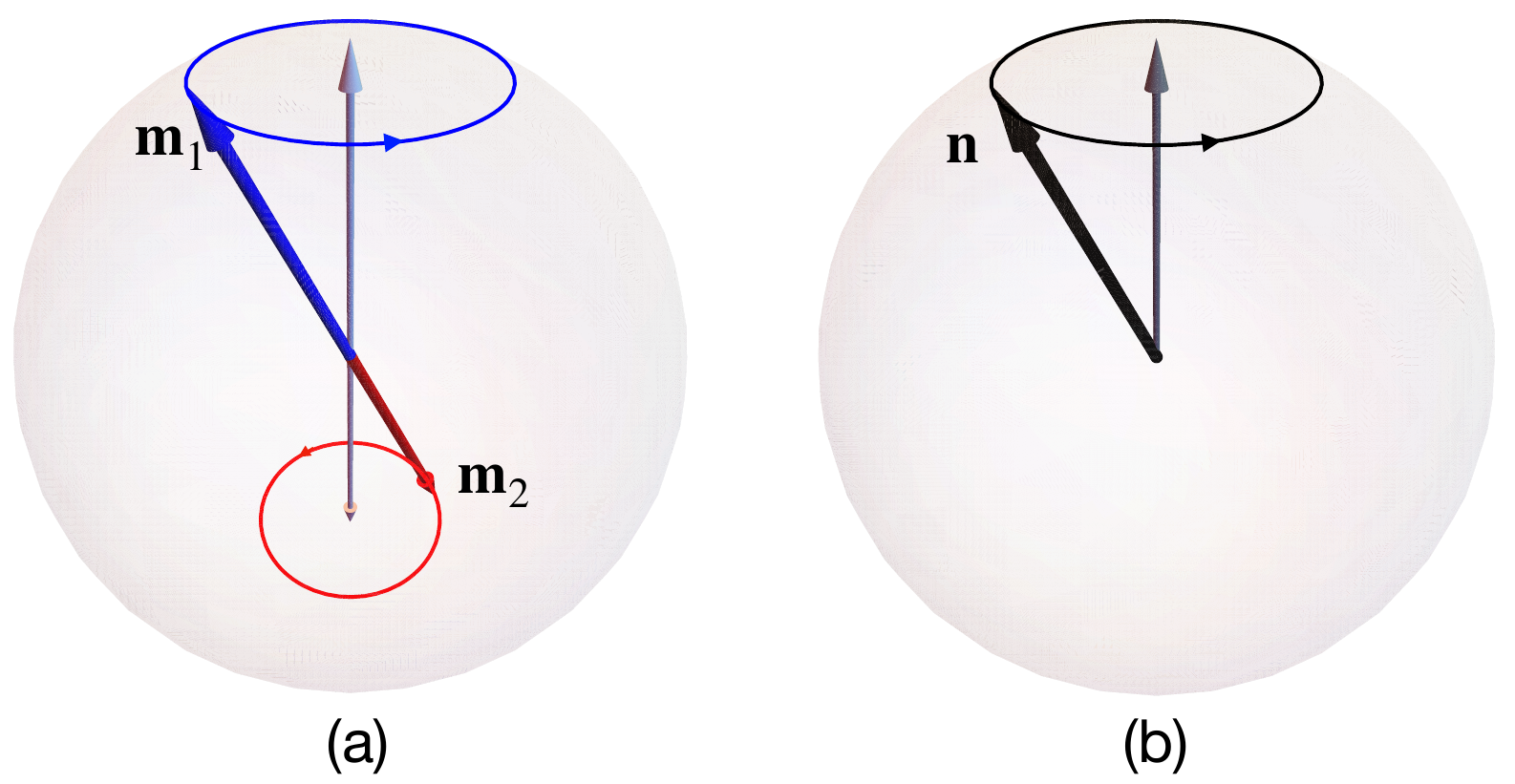}
\caption{Trajectories of the spin order for a well-defined spin projection, $S_{z}$, in a spin wave within the Néel antiferromagnet. (a) The spin directions $\mathbf{m}_a$ ($a = 1, 2$) for right-handed circularly polarized magnons. (b) The Néel vector trajectory around its equilibrium axis, $\mathbf{e}_z$.}
\label{Neel-precession}
\end{figure}

\section{Nonlinear Sigma Model for Non-collinear AFMs}
\label{sec:NoncollinearNLSM}

\begin{figure}[ht]
    \centering
    \includegraphics[width=0.46\textwidth]{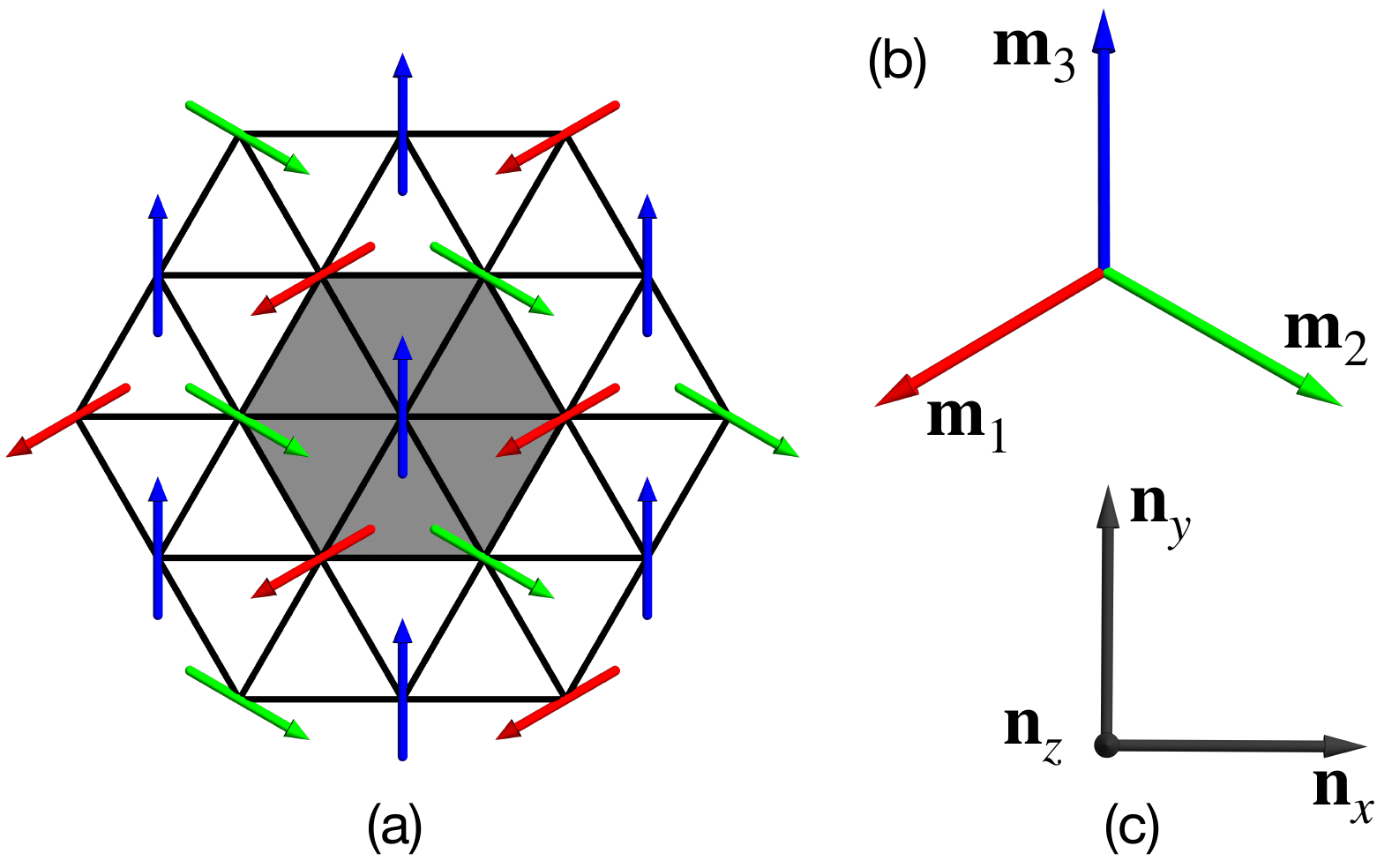}
    \caption{(a) Illustration of the $120^\circ$ order in a three-sublattice antiferromagnet, with the shaded area denoting the magnetic unit cell. (b) Schematic representation of the magnetic order parameters $\mathbf{m}_1$, $\mathbf{m}_2$, and $\mathbf{m}_3$. (c) Depiction of the corresponding spin frame vectors, labeled as $\mathbf{n}_x$, $\mathbf{n}_y$, and $\mathbf{n}_z$.}
    \label{fig:three_sublattice}
\end{figure}

We now turn our attention to non-collinear antiferromagnets, focusing specifically on the triangular Heisenberg antiferromagnet, which exhibits magnetic order characterized by three distinct sublattices. In this system, each sublattice spin is defined as 
\begin{equation}
\mathbf{S}_a = S\, \mathbf{m}_a,\qquad a=1,2,3,
\end{equation}
where $S$ is the spin length and $\mathbf{m}_a$ is a unit vector along the direction of the sublattice magnetization. In the ground state, the three spins within each triangle satisfy
\begin{equation}
\mathbf{S}_1+\mathbf{S}_2+\mathbf{S}_3=0,
\end{equation}
implying that they are coplanar and mutually oriented at $120^\circ$, as illustrated in Fig.~\ref{fig:three_sublattice}(a). In this configuration, the magnetic order parameter may be viewed as a rigid body formed by the unit vectors $\mathbf{m}_1$, $\mathbf{m}_2$, and $\mathbf{m}_3$, which can be rotated collectively without altering the internal structure. Since the overall orientation of a rigid body is characterized by an $\mathrm{SO(3)}$ matrix, the spin order parameter is naturally represented as a matrix of rotations. In fact, Dombre and Read~\cite{dombre1989nonlinear} developed a nonlinear sigma model formulation of the triangular-lattice antiferromagnet based on rotation matrices.

An alternative, but equivalent, formulation was recently introduced by two of us~\cite{pradenas2024spin}, where the order parameter is expressed in terms of an orthogonal \emph{spin frame}. In this approach, the uniform magnetization is defined as,
\begin{equation}
    \mathbf{m}  = \mathbf{m}_1+\mathbf{m}_2+\mathbf{m}_3,
\end{equation}
 while two staggered vectors are defined as
\begin{equation}
\begin{aligned}
\mathbf{n}_x & = \frac{1}{\sqrt{3}} \left( \mathbf{m}_2-\mathbf{m}_1 \right),\\[1mm]
\mathbf{n}_y & = \frac{1}{3}\left( 2\mathbf{m}_3-\mathbf{m}_2-\mathbf{m}_1 \right),
\end{aligned}
\end{equation}
and the vector spin chirality is given by
\begin{equation}
\mathbf{n}_z = \frac{2}{3\sqrt{3}} \Bigl( \mathbf{m}_1 \times \mathbf{m}_2 + \mathbf{m}_2 \times \mathbf{m}_3 + \mathbf{m}_3 \times \mathbf{m}_1 \Bigl).
\end{equation}
In the ground state the uniform magnetization vanishes, $\mathbf{m}=0$, and the spin-frame vectors $\mathbf{n}_x$, $\mathbf{n}_y$, and $\mathbf{n}_z$ form an orthonormal set:
\begin{equation}
\mathbf{n}_a \cdot \mathbf{n}_b = \delta_{ab}, \quad \mathbf{n}_a \times \mathbf{n}_b = \epsilon_{abc}\,\mathbf{n}_c,
\end{equation}
with the Latin indices $a,b,c$ assume the values $x$, $y$, and $z$. The set of spin frame vectors, $\{\mathbf{n}_x, \mathbf{n}_y, \mathbf{n}_z\}$, and the fixed global frame, $\{\mathbf{e}_x, \mathbf{e}_y, \mathbf{e}_z\}$, are connected by the Dombre-Read rotation matrix $O$~\cite{dombre1989nonlinear} as
\begin{equation}
\mathbf{n}_b = \mathbf{e}_a\, O_{ab}, \quad \mathbf{e}_a = O_{ab}\, \mathbf{n}_b .
\end{equation}

As in the conventional two-sublattice antiferromagnets, although the uniform magnetization $\mathbf{m}$ vanishes in the ground state, it becomes nonzero as the spin frame vectors evolve in time. In the low-energy limit, this behavior is captured by
\begin{equation}\label{magnetization-dynamics-3AFM}
\mathbf{m}= \frac{1}{2}\,\chi\, \mathcal{S}\,\mathbf{n}_a \times \partial_t \mathbf{n}_a,
\end{equation}
where sum over repeated indices is assumed, $\chi$ is the magnetic susceptibility, and $\mathcal{S} = S/A_{\mathrm{uc}}$ denotes the spin density, with $S$ being the total spin per magnetic unit cell.

The continuum description of the antiferromagnet in the spin frame is governed by the Lagrangian
\begin{equation} \label{spin-frame-triangular-AFM}
\begin{split}
\mathcal{L} &= \frac{\rho}{4}\,\partial_{t} \mathbf{n}_{a} \cdot \partial_{t} \mathbf{n}_{a} -\frac{\mu}{2}\,\partial_{i} \mathbf{n}_{j} \cdot \partial_{i} \mathbf{n}_{j} \\[1mm]
&\quad -\frac{\lambda}{2}\,\epsilon_{ij}\, \mathbf{n}_z \cdot \Bigl(\partial_i \mathbf{n}_z \times \partial_j \mathbf{n}_z\Bigr).
\end{split}
\end{equation}
Here, the Greek indices $i,j$ take the values $x$ and $y$, while the Latin index $a$ runs over $x$, $y$, and $z$. The parameters $\rho=\chi \mathcal{S}^{2}$ and $\mu$ describe the inertial and exchange stiffness properties of the antiferromagnet, respectively. The final term, which is topological in nature and weighted by $\lambda$, reduces to a boundary contribution in the bulk and is responsible for the emergence of localized helical edge modes~\cite{pradenas2024helical}.  In what follows, we ignore this term for the bulk analysis.

Each vector $\mathbf{n}_a$ can be expressed in terms of small angles of rotations around the ground state~\cite{pradenas2024spin}:
\begin{equation}
\label{eq:small-angles}
\mathbf{n}_a 
= \mathbf{n}_a^{(0)} + \boldsymbol{\phi} \times \mathbf{n}_a^{(0)}+ \mathcal{O}(\boldsymbol{\phi}^2),
\end{equation}
where $\boldsymbol{\phi} = (\phi_x, \phi_y, \phi_z)$ represents small rotation angles around the spin frame vectors in the ground state, $\mathbf{n}_x^{(0)}, \mathbf{n}_y^{(0)}, \mathbf{n}_z^{(0)}$. These angles parametrize to the Goldstone modes arising from the spontaneous breaking of spin-rotation symmetry.

At low energies, the Lagrangian in Eq.~\eqref{spin-frame-triangular-AFM} can be expanded in gradients of the fields $\boldsymbol{\phi}$. Retaining only the quadratic terms yields~\cite{dombre1989nonlinear, dasgupta2020theory, li2021spin}:

\begin{equation}
\label{L-density-triangular-AFM}
\begin{aligned}
\mathcal{L} = \frac{\rho}{2} \Big( \partial_{t}\phi_x^2 &+ \partial_{t}\phi_y^2 
+ \partial_{t}\phi_z^2 \Big) \\
&- \frac{\mu}{2} \Big( \nabla \phi_x^2 + \nabla \phi_y^2  + 2 \nabla \phi_z^2 \Big).
\end{aligned}
\end{equation}
Where the gradient operator is defined as $\nabla = (\partial_x, \partial_y)^T$. This low-energy effective theory predicts three magnon branches, each with distinct propagation speeds. Among these, two branches ($c_{\text{I}}$ and $c_{\text{II}}$) are degenerate, while the third branch ($c_{\text{III}}$) has a higher speed:
\begin{equation}
c_{\text{I,\,II}}
=
\sqrt{\frac{\mu}{\rho}},
\quad
c_{\text{III}}
=
\sqrt{\frac{2\mu}{\rho}}.
\end{equation}

\subsection{Apparent Excess of Symmetries}
\label{subsec:excess of symmetries}

The Lagrangian~\eqref{L-density-triangular-AFM} exhibits three approximate shift symmetries, $\phi_a \rightarrow \phi_a + \alpha_a$ for $a = x, y, z$. These symmetries lead to the Noether currents:
\begin{equation}
j^t_a = \rho\partial_t \phi_a, \quad j^i_{x,y} = -\mu \partial_i \phi_{x,y}, \quad j^i_z = -2\mu \partial_i \phi_z,
\end{equation}
where $a = x, y, z$ and the Latin indices run over $x, y$. The conservation of these currents corresponds directly to the first-order equations of motion.

Additionally, there is an $\mathrm{SO}(2)$ symmetry acting on the field doublet $(\phi_x,\phi_y)$ via the transformation
\begin{equation} \label{eq:SO(2)_triangular}
    \phi_a \rightarrow \phi_a + \beta \epsilon_{ab} \phi_b,
\end{equation}
where $\epsilon_{ab}$ is the two-dimensional Levi-Civita symbol. This transformation gives rise to a conserved current:
\begin{align} 
\label{SO(2)-densities} 
j^t &= \rho \epsilon_{ab} \phi_a \partial_t \phi_b, \quad
j^i = -\mu \epsilon_{ab} \phi_a \partial_{i} \phi_b, 
\end{align}
for $a=x,y$ only. Like the Néel antiferromagnet, these systems exhibit approximate shift symmetries for each magnon branch and feature a single residual $\mathrm{SO}(2)$ symmetry. 

However, unlike in the Néel case where the $\mathrm{SO}(2)$ symmetry can be directly traced to a residual symmetry of the spontaneously broken $\mathrm{SO}(3)$, the situation here is more subtle. In principle, in the triangular-lattice AFM, the non-collinear magnetic order fully breaks $\mathrm{SO}(3)$, leaving no obvious residual symmetry. This raises a fundamental question: \emph{Why does the symmetry count remain comparatively high, and how is the $\mathrm{SO}(2)$ symmetry associated with the degenerate $\phi_x$ and $\phi_y$ magnon modes formally defined and preserved?} 

A plausible explanation is that the order parameter in non-collinear antiferromagnets admits a larger set of transformations than $\mathrm{SO}(3)$. This extended symmetry could account for the persistence of the effective $\mathrm{SO}(2)$ (Eq.~\eqref{eq:SO(2)_triangular}) symmetry in the low-energy spectrum.

\section{The Chiral Symmetry}
\label{sec:Chiral}
In three-dimensional space, an $\mathrm{SO(3)}$ matrix—representing a rigid body's orientation—can be specified in two distinct but equivalent ways, each involving three angles. In one approach, the orientation is achieved through a sequence of extrinsic rotations about axes fixed in the global coordinate system; alternatively, it can be achieved via intrinsic rotations—commonly known as Euler rotations—performed about axes attached to the body (or its principal axes). In our formulation, we consider two right-handed, orthonormal frames. The first is the fixed, global coordinate system with basis vectors $\{\mathbf{e}_x, \mathbf{e}_y, \mathbf{e}_z\}$, and the second is a body-fixed frame with basis vectors $\{\mathbf{n}_x, \mathbf{n}_y, \mathbf{n}_z\}$. The relative orientation between these frames is fully encoded by the rotation matrix $O$, which can be decomposed as
\begin{equation} \label{eq:E-N-decomposition}
O = E^T N,
\end{equation}
where
\begin{equation}
E = \begin{bmatrix} \mathbf{e}_x & \mathbf{e}_y & \mathbf{e}_z \end{bmatrix}, \quad 
N = \begin{bmatrix} \mathbf{n}_x & \mathbf{n}_y & \mathbf{n}_z \end{bmatrix}.
\end{equation}
Here, $E$ and $N$ are $3 \times 3$ matrices whose columns correspond to the basis vectors of the global and body-fixed frames, respectively. In this representation, each element $O_{ab}$ is given by the inner product:
\begin{equation}
O_{ab} = \mathbf{e}_a \cdot \mathbf{n}_b,
\end{equation}
so that the $b$th column of $O$ contains the components of $\mathbf{n}_b$ in the global frame, while the $a$th row provides the components of $\mathbf{e}_a$ in the body frame, respectively. That is,
\begin{equation}
(\mathbf{n}_b)_a = \mathbf{e}_a \cdot \mathbf{n}_b, \quad (\mathbf{e}_a)_b = \mathbf{e}_a \cdot \mathbf{n}_b.
\end{equation}
Next, we examine how rotations relative to the global and body frames act on $O$. An extrinsic rotation, defined relative to the global frame and represented by a rotation matrix $L \in \mathrm{SO(3)}$, transforms the components of the body-fixed vectors as
\begin{equation} \label{eq:L_n_ba}
(\mathbf{n}_b)_a \mapsto L_{ac} \, (\mathbf{n}_b)_c,
\end{equation}
or equivalently, the full rotation matrix transforms as
\begin{equation}
O_{ab} \mapsto L_{ac} O_{cb},\quad  \textrm{or}\quad O \mapsto LO.
\end{equation}

Thus, extrinsic rotations are implemented by left-multiplication of $O$. Conversely, an intrinsic rotation, defined in the body-fixed frame and represented by $R \in \mathrm{SO(3)}$, transforms the components of the global basis vectors as
\begin{equation}
(\mathbf{e}_a)_b \mapsto R_{bc}\, (\mathbf{e}_a)_c,
\end{equation}
so that the rotation matrix transforms according to
\begin{equation}
O_{ab} \mapsto O_{ac}R_{cb} = O_{ac}(R^{T})_{cb},\quad \textrm{or}\quad O \mapsto O R^{T}.
\end{equation}
Thus, intrinsic rotations correspond to right-multiplication of $O$. 
Moreover, these two transformations can act simultaneously, and since left and right multiplications act on different sides of $O$, extrinsic and intrinsic rotations commute. Combined, these simultaneous transformations define the \emph{chiral group},
\begin{equation} \label{eq:chiral-group}
\mathrm{G} = \mathrm{SO(3)_L} \times \mathrm{SO(3)_R},
\end{equation}
where $\mathrm{SO(3)_L}$ corresponds to extrinsic rotations and $\mathrm{SO(3)_R}$ to intrinsic rotations, as hinted by the decomposition of $O$ into the $E$ and $N$ matrices containing the global and body axes, respectively

Therefore, in a non-collinear antiferromagnet the transformations acting on the spin order parameter are not merely ordinary rotations in three-dimensional space; they form a slightly more complex group structure. Also, we choose to retain the name \emph{chiral} because of its close resemblance to the chiral group in particle physics, $\mathrm{SU(2)_L}\times \mathrm{SU(2)_R}$—even though there is no a physical chiral object here, as in contrast to QCD where the left and right $\mathrm{SU}(2)$ groups act on left- and right-handed fermions of the chiral condensate, respectively.

Since the Lie algebra of a direct product group is the direct sum of the Lie algebras of its factors, the Lie algebra of the chiral group is given by
\begin{equation}\label{eq:chiral-algebra}
\mathfrak{so}(3)_{\mathrm{L}} \oplus \mathfrak{so}(3)_{\mathrm{R}}.
\end{equation}
Reflecting the two distinct ways of rotating the spin order orientation, and forming a six dimensional algebra.

The chiral group (\ref{eq:chiral-group}) provides a concise and unified description of the transformations that incorporate both extrinsic and intrinsic rotations acting on the spin order. Interestingly, the resulting symmetry group is highly reminiscent of the principal chiral model in particle physics, which is used to describe the dynamics of mesons, such as pions~\cite{Tong:Gauge}.

\section{O(4)-Symmetric NLSM Perspective} 
\label{sec:O(4)-formulation}

Much like the collinear Néel antiferromagnet, which admits an $\mathrm{O}(3)$ nonlinear sigma model description, a non-collinear AFM on a triangular lattice can be cast in an $\mathrm{O}(4)$ model. Here, rather than describing the spin order through a rotation matrix $O$ or a spin frame, it is more natural to introduce an order parameter defined as a unit vector $q \in S^3$ (with antipodal points identified). In this formulation, the Lie algebra of the chiral group acts directly on vectors, avoiding the less natural two-sided matrix multiplication required in the previous representation (see Eq.~\eqref{eq:chiral-group}). This approach naturally exploits the isomorphism
\begin{equation} \label{eq:isomorphism}
\mathfrak{so}(4) \cong \mathfrak{so}(3)_L \oplus \mathfrak{so}(3)_R,
\end{equation}
and has the added advantage of treating all orientations of the spin order equally on the three-sphere, reflecting the close relationship between $\mathrm{SO}(3)$ and its double cover, $\mathrm{SU}(2)$.

By introducing the four-component unit vector $q$ as a set of coefficients for the $\mathrm{SU}(2)$ matrix
\begin{equation}
    U(q) = q_0 \mathbb{I} - i q_1 \sigma_1 - i q_2 \sigma_2 - i q_3 \sigma_3,
\end{equation}
where $\sigma_a$ are the Pauli matrices, and using that the corresponding rotation matrix can be parametrized as
\begin{equation} \label{eq:Rmatrix-to-q}
    O_{ab} = \left(2 q_0^2 - 1\right) \delta_{ab} + 2 q_a q_b - 2 q_0 \epsilon_{abc} q_c,
\end{equation}
where $a,b,c = 1,2,3$, $\delta_{ab}$ is the Kronecker delta, and $\epsilon_{abc}$ is the Levi-Civita symbol. The theory describing the triangular-lattice AFM described by Eq.~(\ref{spin-frame-triangular-AFM}), takes the NLSM form:
\begin{align} 
\label{eq:triangular-NLSM}
\mathcal{L}
&=
2\rho \Bigl[
(e_1\!\cdot\!\partial_{t}{q})^2 + (e_2\!\cdot\!\partial_{t}{q})^2 + (e_3\!\cdot\!\partial_{t}{q})^2
\Bigr]
\nonumber \\  
&\quad
-\,2\mu \Bigl[
(e_1\!\cdot\!\nabla q)^2 + (e_2\!\cdot\!\nabla q)^2 + 2\,(e_3\!\cdot\!\nabla q)^2
\Bigr].
\end{align}
Here, where the gradient operator is defined as $\nabla = (\partial_x, \partial_y)^T$. Additional orthonormal tangent vectors $e_a$ (with $a=1,2,3$) describe the local deviations around $q$. Since these vectors are orthogonal to $q$, together they form a complete basis in four-dimensional space.

The more symmetric $\mathrm{O}(4)$-NLSM formulation naturally reveals the three distinct magnon branches that characterize the low-energy spectrum of the antiferromagnet on a triangular lattice. Just as in the Néel antiferromagnet, where magnons are understood as small deviations of the unit vector $\mathbf{n}$ from a reference state, here magnons are interpreted as small deviations of the four-dimensional vector $q$ from a reference configuration, with their contributions captured by the projection onto the tangent-space vectors.

For a universal symmetry-breaking analysis and for pedagogical purposes, we further reduce the system to the fully symmetric $\mathrm{O(4)}$ \textit{principal chiral model} (PCM) in two spatial dimensions:
\begin{equation}\label{eq:O(4)-Symmetric NLSM}
\mathcal{L}_{\mathrm{PCM}}
= 2\rho\,(\partial_t q)^2-2\mu\,(\nabla q)^2.
\end{equation}
This Lagrangian is invariant under global $\mathrm{SO}(4)$ rotations acting on $q \in S^3$. Such models capture topological excitations (e.g., skyrmions, baby skyrmions) in non-collinear or non-coplanar AFMs~\cite{batista2018principal} and illuminate how additional internal symmetries arise from the geometrical structure of $S^3$ as we will see in the next section.

We will later revisit the triangular-lattice antiferromagnet described by Eq.~(\ref{eq:triangular-NLSM}), which represents a system with lower symmetry than that of the principal chiral model.

\section{Chiral Symmetry Breaking}
\label{sec:Chiral Symmetry Breaking}

In the $\mathrm{O}(4)$-NLSM formulation, the infinitesimal extrinsic and intrinsic rotations act directly on the spin order parameter $q$, rather than through two-sided matrix multiplication. This simplification follows from the isomorphism in Eq.~\eqref{eq:isomorphism}, and takes the form:
\begin{equation} \label{eq:so(4)-infinitesimal}
dq = \Bigl( d\phi_a \lambda_a + d\psi_a \rho_a \Bigr) q.
\end{equation}
Here, $d\phi_a$ and $d\psi_a$ are infinitesimal rotation angles associated with the global and body-fixed frames, respectively. The matrices $\lambda_a$ generate extrinsic (global) rotations, while $\rho_a$ generate intrinsic (body-fixed) rotations. The explicit forms of these generators are 
\begin{widetext}
\begin{equation} \label{Left-SO(4)}
\lambda_1 = \frac{1}{2}
\begin{pmatrix}
0 & -1 & 0 & 0 \\
1 & 0 & 0 & 0 \\
0 & 0 & 0 & -1 \\
0 & 0 & 1 & 0
\end{pmatrix}, 
\quad
\lambda_2 = \frac{1}{2}
\begin{pmatrix}
0 & 0 & -1 & 0 \\
0 & 0 & 0 & 1 \\
1 & 0 & 0 & 0 \\
0 & -1 & 0 & 0
\end{pmatrix},
\quad
\lambda_3 = \frac{1}{2}
\begin{pmatrix}
0 & 0 & 0 & -1 \\
0 & 0 & -1 & 0 \\
0 & 1 & 0 & 0 \\
1 & 0 & 0 & 0
\end{pmatrix},
\end{equation}
for global rotations, and
\begin{equation} \label{Right-SO(4)}
\rho_1 = \frac{1}{2}
\begin{pmatrix}
0 & -1 & 0 & 0 \\
1 & 0 & 0 & 0 \\
0 & 0 & 0 & 1 \\
0 & 0 & -1 & 0
\end{pmatrix}, 
\quad
\rho_2 = \frac{1}{2}
\begin{pmatrix}
0 & 0 & -1 & 0 \\
0 & 0 & 0 & -1 \\
1 & 0 & 0 & 0 \\
0 & 1 & 0 & 0
\end{pmatrix},
\quad
\rho_3 = \frac{1}{2}
\begin{pmatrix}
0 & 0 & 0 & -1 \\
0 & 0 & 1 & 0 \\
0 & -1 & 0 & 0 \\
1 & 0 & 0 & 0
\end{pmatrix}.
\end{equation}
\end{widetext}
for the generators of body-fixed rotations.

The commutation relations for these generators correspond to two commuting $\mathfrak{so}(3)$ algebras:
\begin{equation} \label{eq:lie-algebra-SO(4)}
    [\lambda_{a},\lambda_{b}] = \epsilon_{abc }\lambda_{c}, \; [\rho_{a},\rho_{b}] = -\epsilon_{abc } \rho_{c}, \; [\lambda_{a},\rho_{b}] = 0.
\end{equation}
These relations define the Lie algebra of the SO(4) group, which describes rotations in four-dimensional space. The resulting algebra is six-dimensional, corresponding to the six independent planes in 4D space where rotations can occur. Notice that the generators of the body-axis rotations, $\rho_a$, satisfy what are sometimes called anomalous commutation relations \cite{biedenharn1984angular, landau2013quantum}.

The same algebra~\eqref{eq:lie-algebra-SO(4)} appears in the quantum mechanical treatment of angular momentum in molecules and more general rigid rotors, where the rotational states of a rigid molecule are typically described in terms of its total angular momentum and its projections onto both globally fixed and the molecule's internal axes (body-fixed axes)~\cite{landau2013quantum}.

The full invariance of the $\mathrm{O(4)}$-NLSM (Eq.~(\ref{eq:O(4)-Symmetric NLSM})) under global transformations of the chiral group $\mathrm{SO(3)_L \times SO(3)_R}$ gives rise to conserved spin currents:
\begin{align}\label{current-global}
j_{\textrm{L},a}^{\alpha} &= 
\frac{\partial \mathcal{L}}{\partial (\partial_{\alpha}q)} 
\cdot \frac{\partial q}{\partial \phi_{a}} 
= 4 \rho \, \partial^{\alpha} q \cdot \lambda_{a}\, q, \\
\label{current-body-axis}
j_{\textrm{R},a}^{\alpha} &= 
\frac{\partial \mathcal{L}}{\partial (\partial_{\alpha}q)} 
\cdot \frac{\partial q}{\partial \psi_{a}} 
= 4 \rho \, \partial^{\alpha} q \cdot \rho_{a} q.
\end{align}
Here, the Greek index $\alpha$ denotes the spacetime components of the current, and we adopt the $(+,-,-)$ metric convention. Furthermore, we define the magnon speed of propagation as $c = \sqrt{\mu/\rho}$. It is important to emphasize that the conservation of these currents, $\partial_{\alpha} j_{(\textrm{L,R}),a}^{\alpha} = 0$, corresponds to the projections of the Landau--Lifshitz equations onto the two sets of axes: the three global axes and the three body-fixed axes.

The time components of these Noether currents represent the projections of the spin density, $\mathcal{S}\mathbf{m}$ (see Eq.~\eqref{magnetization-dynamics-3AFM}), onto the fixed global and body-attached axes:
\begin{align}\label{eq:left-right-currents}
  \sigma_{\textrm{L},a} &= j_{\textrm{L},a}^t = \mathcal{S}\mathbf{m} \cdot \mathbf{e}_{a}, \quad
  \sigma_{\textrm{R},a} = j_{\textrm{R},a}^t = \mathcal{S}\mathbf{m} \cdot \mathbf{n}_{a}.
\end{align}
To explore the breaking of the chiral symmetry by the spin-order ground state, we take, without loss of generality, the ground state to be
\begin{equation}\label{eq:q-ground state}
q_{\text{gs}} = (1, 0, 0, 0).
\end{equation}
This state corresponds to full alignment between the global frame and the spin-frame axes (i.e., $\mathbf{e}_a = \mathbf{n}_a$). In this configuration, both extrinsic and intrinsic rotations are broken, as indicated by
\begin{equation}
   \lambda_{a}\, q_{\text{gs}} \neq 0,\quad  \rho_{a}\, q_{\text{gs}} \neq 0.
\end{equation}

However, the breaking of the generators is not complete but \emph{partial}, since a particular combination of extrinsic and intrinsic rotations leaves the ground state invariant. To isolate the unbroken part of the generators, we define the \emph{vector} and \emph{axial} components of the chiral group as
\begin{equation} \label{eq:def-axial-vector}
    V_a = \lambda_a - \rho_a, \quad A_a = \lambda_a + \rho_a.
\end{equation}
Here, the vector generators satisfy the usual commutation relations of rotations in three dimensional space,
\begin{equation}\label{eq:so(3)-vector}
[V_a, V_b] = \epsilon_{abc} V_c,
\end{equation}
while the commutators between the axial and vector generators are
\begin{equation} \label{eq:axial-vector}
     [A_a, V_b]  = \epsilon_{abc} A_c, \quad [A_a, A_b]  = \epsilon_{abc} V_c.
\end{equation}
Indeed, the ground state of the $\mathrm{O}(4)$ nonlinear sigma model (Eq.~\eqref{eq:O(4)-Symmetric NLSM}) remains invariant under the subalgebra generated by $V_a$. This defines the residual group: 
\begin{equation} \label{eq:residual_SO(3)} 
\mathrm{SO}(3)_{\mathrm{V}} \equiv \mathrm{SO}(3)_{\mathrm{L}} \times \mathrm{SO}(3)_{\mathrm{R}={L}^{-1}}.
\end{equation} 
Transformations in this residual group correspond to conjugations (similarity transformations) of the rotation matrix $O$ by another matrix $L$, explicitly given by
\begin{equation} \label{eq:axis-angle-conjugation}
O \to L O L^{-1}.
\end{equation}
This implies that the residual $\mathrm{SO(3)}_\textrm{V}$ symmetry is realized through simultaneous extrinsic rotations of the global axes and intrinsic rotations of the corresponding body-fixed axes, performed by equal and opposite angles (i.e., $\phi_a = -\psi_a$, as introduced in Eq.~\eqref{eq:so(4)-infinitesimal}).  In contrast, the axial generators are completely broken by the ground state. Since the symmetries of the Lagrangian differ from those of the ground state, we conclude that the ground state of the principal chiral model breaks $\mathrm{SO(3)_{L}\times SO(3)_{R}}$ down to $\mathrm{SO(3)_{V}}$.

\section{The Isospin Quantum Number}
\label{sec:isospin}

The breaking of chiral symmetry gives rise to three massless Goldstone modes in the low-energy theory, each associated with one of the broken axial directions $A_a$. From Eq.~\eqref{eq:axial-vector}, we observe that the axial generators $A_a$ transform as a 3-vector under the unbroken $\mathrm{SO(3)_V}$ subgroup. Therefore, the Goldstone modes also transform as a triplet under $\mathrm{SO(3)_V}$. Around the uniform ground state defined in Eq.~\eqref{eq:q-ground state}, these modes are parametrized by the components $q_1$, $q_2$, and $q_3$, as
\begin{equation}\label{eq:low-energy-state}
q = (q_{0},\, \mathbf{q}),
\end{equation}
where $q_{0}=\sqrt{1-\mathbf{q}^2}$ and $\mathbf{q} = (q_1, q_2, q_3)$. The residual $\mathrm{SO(3)_V}$ transformations consist of rotations that act exclusively on the components $q_1$, $q_2$, and $q_3$, while preserving both $q_1^2 + q_2^2 + q_3^2$ and $q_0^2$ independently. This invariance is manifestly reflected in the trace of the spin order parameter $O$, which remains unchanged under residual group transformations~\eqref{eq:residual_SO(3)}:
\begin{equation}\label{eq:TRR}
    \mathrm{Tr}[P O P^{-1}] = \mathrm{Tr}[O] = 4q_0^2 - 1.
\end{equation}
Here, $P$ is an arbitrary three-dimensional rotation matrix, and we have used the relation~\eqref{eq:Rmatrix-to-q}. This reflects the invariance of the Goldstone mode amplitude under the residual $\mathrm{SO(3)_V}$ transformations.

For this residual symmetry, we define a corresponding Noether current. In particular, using Eqs. \eqref{current-global} and ~\eqref{eq:def-axial-vector}, the conserved \emph{vector} current is given by
\begin{equation} \label{eq:isospin_current}
    j_{\textrm{V},a}^{\alpha} = 4 \rho \, \partial^{\alpha} q \cdot V_{a} q.
\end{equation}

Its associated charge density is
\begin{equation} \label{eq:isospin_charge density}
    \sigma_{\textrm{V},a}=j_{\textrm{V},a}^{t} = 4 \rho  \epsilon_{abc} \, q_{b} \, \partial_{t}q_{c}.
\end{equation}

Furthermore, combining Eq.~\eqref{eq:left-right-currents} and~\eqref{eq:def-axial-vector}, this charge density can be interpreted as the difference between the projections of the spin density onto the global and body-fixed axes:
\begin{equation}
    \sigma_{\textrm{V},a} = \mathcal{S}\,\mathbf{m}\cdot\mathbf{e}_{a} - \mathcal{S}\,\mathbf{m}\cdot\mathbf{n}_{a}.
\end{equation}

We refer to Eq.~\eqref{eq:isospin_current} as the \emph{isospin} (short for ``isotopic spin") current because the Lie algebra of the global charges (i.e., $V_{a}$ generators) is identical to that of ordinary spin. Although isospin is not related to intrinsic angular momentum, it emerges as a conserved quantity rooted in the symmetry properties of non-collinear antiferromagnets. These charges, arising from the residual $\mathrm{SO(3)_V}$, label an internal degree of freedom carried by magnons. A similar situation occurs in particle physics, where the spontaneous breaking of chiral symmetry by the QCD vacuum leads to pions as Goldstone bosons, with the corresponding isospin charge.

To see how the isospin charge can be used to label different magnon states—and, in particular, how it can be promoted to a quantum label—we consider states near the ground state~\eqref{eq:q-ground state}. In this regime, where $|\mathbf{q}| \ll 1$, the order parameter~\eqref{eq:low-energy-state} takes the approximate form $q = (q_0, \mathbf{q}) \approx (1, \mathbf{q})$, the chiral principal model Lagrangian~\eqref{eq:O(4)-Symmetric NLSM} reduces to:
\begin{align}\label{eq:Non-Collinear-NLSM}
\mathcal{L} &= 2\rho (\partial_t \mathbf{q})^2 - 2\mu (\nabla \mathbf{q})^2 + \ldots
\end{align}
In the small-amplitude limit, the rotation matrix $O = e^{\phi_x L_x} e^{\phi_y L_y} e^{\phi_z L_z}$ can be approximated by $O_{ij} \approx \delta_{ij} - \epsilon_{ijk} \phi_k$. Comparing this with Eq.~\eqref{eq:Rmatrix-to-q}, we recognize that the field triplet
\begin{equation}\label{eq:quaternions-to-small-angles}
\mathbf{q} = (q_1, q_2, q_3) = \frac{1}{2}(\phi_x, \phi_y, \phi_z)
\end{equation}
parametrizes the three Goldstone modes in the small-amplitude regime, as specified in Eq.~\eqref{eq:small-angles}. Consequently, the chiral principal model Lagrangian becomes
\begin{equation} \label{eq:Non-Collinear-NLSM-phi}
\mathcal{L} = \frac{\rho}{2} (\partial_t \boldsymbol{\phi})^2 - \frac{\mu}{2} (\nabla \boldsymbol{\phi})^2 + \ldots
\end{equation}
In contrast to the triangular Heisenberg antiferromagnet, where only two of the branches are degenerate, the principal chiral model features three Goldstone branches that all propagate with the same velocity, $c_{\mathrm{I,II,III}} = \sqrt{\mu/\rho}$. As a consequence, the residual symmetry is not $\mathrm{SO(2)}$ but a higher $\mathrm{SO(3)_V}$. Under a global $\mathrm{SO(3)_V}$ transformation, the Goldstone fields transform as
\begin{equation} \label{eq:SO(3)_V}
\phi_{a} \rightarrow \phi_{a} + \epsilon_{abc}\,\beta_b\,\phi_{c},
\end{equation}
where $\beta_b$ are small constant parameters and $\epsilon_{abc}$ is the totally antisymmetric tensor. These transformations leave the quadratic Lagrangian~\eqref{eq:Non-Collinear-NLSM-phi} invariant. The current components in Eq.~\eqref{eq:isospin_current} become
\begin{equation} \label{eq:linearized_isospin_current}
j_{\textrm{V},a}^{t} = \rho \epsilon_{abc}\phi_{b}\partial_t\phi_{c}, \quad j_{\textrm{V},a}^{i} = \mu \epsilon_{abc}\phi_{b}\partial^{i}\phi_{c}.
\end{equation}

The above currents are the $\mathrm{SO(3)_V}$ analogue of the $\mathrm{SO(2)}$ charge density found in collinear Heisenberg antiferromagnets~(\ref{SO(2)-densities}).

To get the isospin quantum label, we quantize the three fields $\phi_a$ by imposing the canonical commutator
\begin{equation}
\bigl[\phi_{a}(x),\,\pi_{b}(y)\bigr]
=
i\hbar\,\delta_{ab}\delta^{2}\bigl(\mathbf{x}-\mathbf{y}\bigr),
\end{equation}
where the canonical momentum is defined in the standard way as
$\pi_{a}(x) = \partial\mathcal{L}/\partial (\partial_{t}\phi_{a}(x)) $. To separate left- and right-handed polarization, we define
\begin{equation}
\phi_{\pm}
=
\frac{1}{\sqrt{2}}
\bigl(
\phi_{x}
\pm
i\phi_{y}
\bigr),
\quad
\phi_{0}
=
\phi_{z}.
\end{equation}
To quantize the fields, we first introduce bosonic operators—namely, the annihilation operators $a_{\mathbf{k}}$, $b_{\mathbf{k}}$, and $c_{\mathbf{k}}$, along with their corresponding creation operators $a_{\mathbf{k}}^\dagger$, $b_{\mathbf{k}}^\dagger$, and $c_{\mathbf{k}}^\dagger$—which satisfy the standard commutation relations (e.g., $[a_{\mathbf{k}},a_{\mathbf{k}'}^\dagger]=\delta_{\mathbf{k},\mathbf{k}'}$, etc.). In terms of these operators, the fields are expanded in plane-wave modes with well-defined frequency, momentum, and polarization as follows:
\begin{equation}\label{canonical_quantization}
\begin{split}
\phi_{+}(x) &= \sqrt{\frac{\hbar}{\rho A}} \sum_{\mathbf{k}} \frac{1}{\sqrt{2\omega }} \Bigl( b_{\mathbf{k}}\, e^{-i kx} + a_{\mathbf{k}}^\dagger\, e^{i kx} \Bigr),\\[6pt]
\phi_{-}(x) &= \sqrt{\frac{\hbar}{\rho A}} \sum_{\mathbf{k}} \frac{1}{\sqrt{2\omega}} \Bigl( a_{\mathbf{k}}\, e^{-i kx} + b_{\mathbf{k}}^\dagger\, e^{i kx} \Bigr),\\[6pt]
\phi_{0}(x) &= \sqrt{\frac{\hbar}{\rho A}} \sum_{\mathbf{k}} \frac{1}{\sqrt{2\omega}} \Bigl( c_{\mathbf{k}}\, e^{-i kx} + c_{\mathbf{k}}^\dagger\, e^{i kx} \Bigr),
\end{split}
\end{equation}
where $kx=\omega t-\mathbf{k}\cdot\mathbf{x}$, $A$ is the sample area, and the dispersion relation is given by $\omega=\sqrt{\rho/\mathcal{J}}\,|\mathbf{k}|$. 

The polarization is directly related to the third component of the isospin density, as given in Eq.~\eqref{eq:linearized_isospin_current}. The corresponding global isospin charge is obtained by integrating this density over all space. In terms of the quantized fields, becomes
\begin{equation}
   Q_{\textrm{V},3} = \mathopen{:} \int d^2 x \; \sigma_{\textrm{V},3} \; \mathopen{:} = \hbar \sum_{\mathbf{k}} 
\left( a_{\mathbf{k}}^{\dagger} a_{\mathbf{k}}-b_{\mathbf{k}}^{\dagger} b_{\mathbf{k}}  \right).\\   
 \end{equation}

Here, the symbol $\mathopen{:}\mathclose{:}$ denotes normal ordering of operators. Defining the total number operators for the quanta $\phi_{+}$ and $\phi_{-}$ as $N_{a}$ and $N_{b}$, respectively. The isospin charge becomes
\begin{equation}
     Q_{\textrm{V},3} = \hbar (N_{a} -N_{b}) .
\end{equation}
Therefore, magnons come in three different charge states defined by their isospin projection $I_{3}$:
\begin{align}
   & I_{3}(\phi_{+} ) = +\hbar,\; I_{3}(\phi_{0} ) = 0, \;  I_{3}(\phi_{-} ) = -\hbar. 
\end{align}
This represents one of the central results of this paper. Here, we have arbitrarily chosen the third axis, taking advantage of the full rotational $\mathrm{SO(3)_V}$ symmetry of the principal chiral model. In contrast, for an antiferromagnet on a triangular lattice, the axis choice is not arbitrary but is physically restricted to be associated to the direction orthogonal to the spin-plane.

\subsection{Isospin in a Heisenberg AFM in a triangular lattice}

Even though the ground states of both the principal chiral model and the antiferromagnet on a triangular lattice can, without loss of generality, be written as
\begin{equation}\label{eq:matrix-gs}
O = E^{T} N = \mathbb{I},
\end{equation}
and are fully symmetric under the combined transformations of the residual vector group $\mathrm{SO(3)_V}$ defined by $O \rightarrow P O P^{-1}$ the two systems exhibit \emph{subtle yet fundamental} differences in their symmetry structure.  

Formally, the Lagrangian describing the triangular lattice antiferromagnet, Eq.~\eqref{eq:triangular-NLSM}, differs from the fully $\mathrm{SO(4)}$-symmetric principal chiral model, Eq.~\eqref{eq:O(4)-Symmetric NLSM}, by a single symmetry-breaking term,
\begin{equation}
\mathcal{L} = \mathcal{L}_{\mathrm{PCM}} - 2\mu \bigl(e_3 \cdot \nabla q\bigr)^2,
\end{equation}
where $\mathcal{L}_{\mathrm{PCM}}$ is the Lagrangian of the fully symmetric principal chiral model.  Because this additional term is invariant only under rotations about the out‑of‑plane vector $\mathbf{n}_z$, it reduces the body‑rotation symmetry from $\mathrm{SO(3)_R}$ to its subgroup $\mathrm{SO(2)_{R,3}}$. Therefore, the full symmetry group of the antiferromagnet on a triangular lattice is 
\begin{equation}
   \mathrm{SO(3)_L} \times \mathrm{SO(2)_{R,3}},
\end{equation}
where $\mathrm{SO(3)_L}$ acts by global rotations and $\mathrm{SO(2)_{R,3}}$ consists of intrinsic rotations about $\mathbf{n}_z$, the axis perpendicular to the spin plane.

In the principal chiral model, the residual $\mathrm{SO(3)_V}$ symmetry remains manifest in the Goldstone modes because all three spin-wave branches are degenerate. In contrast, for the triangular lattice antiferromagnet only two of the three branches are degenerate, with $c_{\textrm{I,II}} = c_{\textrm{III}}/\sqrt{2}$. As a consequence, although the ground state retains full $\mathrm{SO(3)_V}$ invariance, the magnon excitations in the triangular lattice antiferromagnet do not. Instead, the residual symmetry is reduced to the smaller subgroup 
$\mathrm{SO(2)_{V,3}} \subset \mathrm{SO(3)_V}$, as hinted in Sec.~\ref{subsec:excess of symmetries}. 
Here, the subscript 3 indicates that the residual isospin symmetry for the triangular antiferromagnet consists solely of matrix conjugation by $R_z$, with the $z$-axis defined as perpendicular to the spin plane in the ground state.

Solutions that are $\mathrm{SO(2)_{V,3}}$ symmetric correspond to circularly polarized magnons. For instance, magnons with right-handed circular polarization and finite amplitude $\Theta_0$ around the ground state~\eqref{eq:matrix-gs} have the following spin order parameter:
\begin{equation} \label{eq:fin_amp_circ_pol}
 O = R_z(\Phi)\, R_x(\Theta_{0})\, R_z^{-1}(\Phi),
\end{equation}
where $\Phi = \omega t - \mathbf{k} \cdot \mathbf{x}$. The resulting matrix describes a rotation by $\Theta_0$ about the axis 
$ \mathbf{n} = R_z(\Phi)(1,0,0)^T = (\cos \Phi, \sin \Phi, 0)^T$. Alternatively, the above rotation matrix is equivalent to an Euler rotation in the (3-1-3) convention with Euler angles $(\phi, \theta, \psi) = (\Phi, \Theta_0, -\Phi)$.

\begin{figure}[ht]
    \centering
    \includegraphics[width=0.46\textwidth]{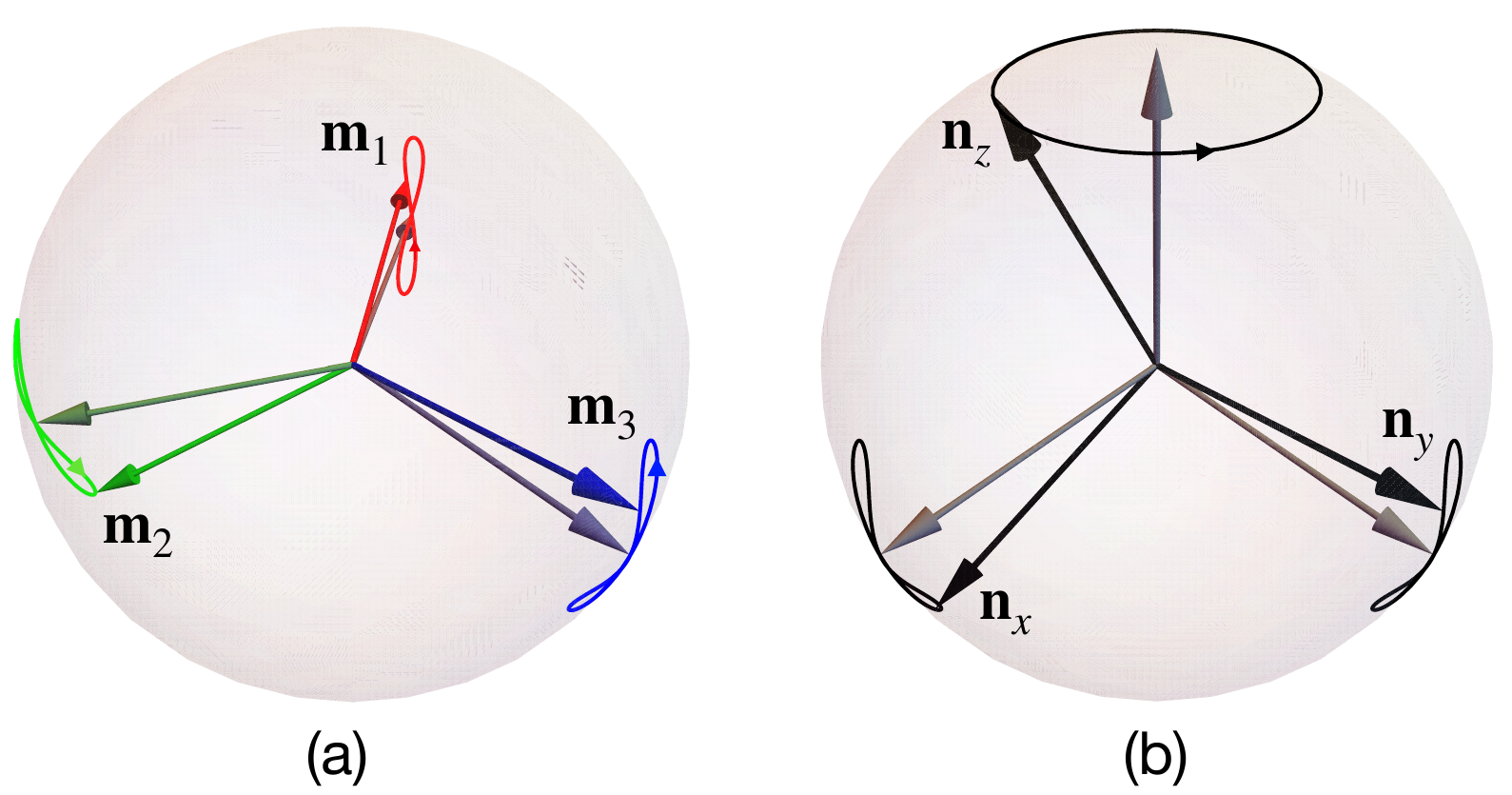}
    \caption{Trajectory of the spin order for a well-defined isospin.
Panel (a) shows the trajectories of the spin directions $\mathbf{m}_a$, with $a=1,2,3$, for right-handed circularly polarized magnons. Panel (b) displays the corresponding trajectories of the spin-frame vectors. Notably, the $x$ and $y$ components of $\mathbf{n}_x$ and $\mathbf{n}_y$ oscillate at twice the frequency of the other components, producing Lissajous-like patterns on the unit sphere. In contrast, $\mathbf{n}_z$ undergoes uniform precession around its equilibrium axis, $\mathbf{e}_z$, at frequency $\omega$.}
    \label{fig:wobbling_motion}
\end{figure}

The resulting dynamics of the spin order Eq.\eqref{eq:fin_amp_circ_pol} correspond to a periodic ``wobbling'' motion of the spin plane around the ground state, see Fig.\eqref{fig:wobbling_motion}—akin to a coin wobbling on a table, but without an overall rotation of the coin. In this picture, the angular displacement $\Theta_0$ acts as the coin’s tilt, while the phase $\Phi$ causes the spin plane to oscillate about its equilibrium. Unlike a coin that might roll or spin, the spin order remains anchored to the ground state, with only its orientation undergoing oscillatory deviations.

This finite amplitude oscillatory behavior has a direct impact on the magnon dispersion. In fact, it modifies the dispersion relation for circularly polarized magnons to
\begin{equation} \label{eq:so(3)-exact}
\omega^2 = \Bigl(2 - \cos \Theta_0 \Bigr) c_{\mathrm{I}}^2 |\mathbf{k}|^2,
\end{equation}
where $c_{\mathrm{I}} = \sqrt{\mu / \rho}$. Thus, the effective ``stiffness" is modified by the finite wave amplitude $\Theta_0$.

Furthermore, using Eqs.~\eqref{eq:left-right-currents} and~\eqref{eq:def-axial-vector}, the axial and vector (isospin) charge densities are given by
\begin{equation}
\begin{aligned} \label{eq:A-V-charge-densities}
\sigma_{\textrm{A},3} &= 0, \quad \sigma_{\textrm{V},3} = 4\rho\omega \sin^2\left(\Theta_0/2\right).
\end{aligned}
\end{equation}
For small amplitudes, we recover that $\sigma_{\textrm{V},3}$ is equivalent to the charge density in \eqref{SO(2)-densities} for circularly polarized magnons, where the amplitude is related to the infinitesimal angle of rotations by $\Theta_{0} = \sqrt{\phi_x^2 + \phi_y^2}$.


\section{Discussion}
\label{sec:Discussion}

We have presented a detailed theoretical analysis of the symmetry group in systems described by an $\mathrm{SO(3)}$ order parameter, with particular emphasis on the non-collinear magnetic order of the Heisenberg antiferromagnet on a triangular lattice. Using the principal chiral model as a guiding framework, we have demonstrated that the full symmetry group of this system is, in fact, the chiral group $\mathrm{SO(3)_L \times SO(2)_{R,3}}$—encompassing both extrinsic (space-fixed) and intrinsic (body-fixed) rotations about the axis $\mathbf{n}_{z}$ perpendicular to the spin plane—rather than the simpler $\mathrm{SO(3)}$ symmetry that is commonly assumed.

The chiral group naturally decomposes into a \emph{vector} part and an \emph{axial} part, each consisting of specific combinations of extrinsic and intrinsic rotations. The magnetic order fully breaks the axial part, giving rise to three magnons associated with the three broken axial generators. Meanwhile, the vector subgroup remains unbroken and defines the residual symmetries of the vacuum.

Using this framework, we clarify the internal $\mathrm{SO}(2)$ symmetry responsible for the degeneracy of two spin-wave branches in the triangular-lattice Heisenberg antiferromagnets as the residual $\mathrm{SO(2)_{V,3}}$ symmetry emerges from a particular combination of joint rotations about the global axis $\mathbf{e}_z$ and the local spin-frame axis $\mathbf{n}_z$, inherited from the larger chiral group.

Furthermore, these results reveal a new quantum observable that distinguishes left- and right-circular magnons, going beyond the usual classification by momentum and energy. Classically, these modes are simply oppositely polarized spin waves. Once quantized, however, they form an isospin doublet with total isospin $I=\hbar$, split into projections $I_3=\pm \hbar$. This additional quantum label thus complements the standard momentum and energy labels, reflecting the deeper symmetry structure inherent to non-collinear antiferromagnets.

The control and manipulation of the emergent isospin of magnons opens up exciting prospects for engineering interactions that depend solely on the isospin charge states. For example, a recent study of a fully symmetric $\mathrm{O}(4)$-NLSM model~\cite{zarzuela2024non}––which serves as a conceptual framework for both spin glass and frustrated antiferromagnet systems––revealed a distinctive Hall-like response arising when magnons in different polarization states scatter off various types of three-dimensional topological solitons. Under the lens of our work, these magnons can be understood as distinct isospin charge states that interact with topological solitons via an isospin-dependent coupling.

We surmise that this higher-symmetry framework could be extended to other families of non-collinear antiferromagnets with more intricate symmetry-breaking patterns. In such systems, we anticipate that analogous residual symmetries will play a fundamental role in defining similar quantum numbers and imposing additional conservation laws.

\section*{Acknowledgments} We thank Boris Ivanov, Ibrahima Bah, and especially Hua Chen for a series of stimulating and insightful discussions. We also thank Bertrand Delamotte and Dominique Mouhanna for drawing our attention to Refs.~\cite{PhysRevLett.64.3175, PhysRevLett.70.2483}. This research was supported by the U.S. Department of Energy under Award No. DE-SC0019331 and the U.S. National Science Foundation under Grant No. NSF PHY-2309135.

\bibliography{references}

\begin{widetext}
\appendix

\section{$\mathrm{ SO(3)_{L} \times SO(3)_{R}}$}

Consider two right-handed orthonormal frames: a \emph{global} coordinate system with basis vectors $\{\mathbf{e}_x, \mathbf{e}_y, \mathbf{e}_z\}$, and a \emph{body-attached} frame with basis vectors $\{\mathbf{n}_x, \mathbf{n}_y, \mathbf{n}_z\}$. The relative orientation between these two frames is fully specified by a rotation matrix $O$, which can be decomposed as

\begin{equation}
    O = E^{T} N,
\end{equation}
where 
\begin{equation}\label{eq:def-EN}
    E = \begin{bmatrix} 
    \mathbf{e}_x & \mathbf{e}_y & \mathbf{e}_z 
    \end{bmatrix},
    \quad
    N = \begin{bmatrix} 
    \mathbf{n}_x & \mathbf{n}_y & \mathbf{n}_z 
    \end{bmatrix}.
\end{equation}

Explictly:

\begin{equation}
O =
\begin{bmatrix}
\mathbf{e}_x \cdot \mathbf{n}_x & \mathbf{e}_x \cdot \mathbf{n}_y & \mathbf{e}_x \cdot \mathbf{n}_z \\
\mathbf{e}_y \cdot \mathbf{n}_x & \mathbf{e}_y \cdot \mathbf{n}_y & \mathbf{e}_y \cdot \mathbf{n}_z \\
\mathbf{e}_z \cdot \mathbf{n}_x & \mathbf{e}_z \cdot \mathbf{n}_y & \mathbf{e}_z \cdot \mathbf{n}_z
\end{bmatrix}.
\end{equation}

Now, we assume that the global (lab) frame is defined by the standard Cartesian axes, so that the matrix $E$ is the identity, i.e., $E=I$. In this case, the rotation matrix relating the body frame to the global frame simplifies to
\begin{equation}
O = E^{T} N = I^{T} N = N.
\end{equation}
Now, consider multiplying $R$ on the left by a rotation about the $z$-axis, denoted by
\begin{equation}
R_{z}(\theta) = \begin{bmatrix} 
\cos\theta & -\sin\theta & 0 \\
\sin\theta & \cos\theta  & 0 \\
0          & 0           & 1 
\end{bmatrix}.
\end{equation}
The new rotation matrix is given by
\begin{equation}
O' = R_{z}(\theta) O = R_{z}(\theta) N.
\end{equation}
Since left-multiplication by $R_{z}(\theta)$ corresponds to rotating the \emph{global} coordinate system, this operation effectively re-expresses the body frame axes (originally given by the columns of $N$) in a new, globally rotated frame. In other words, each column of $O'$ is obtained by applying the rotation $R_{z}(\theta)$ to the corresponding body axis:
\begin{equation}
\textbf{Extrinsic rotations:} \quad     O' = \begin{bmatrix}
R_{z}(\theta)\, \mathbf{n}_x & R_{z}(\theta)\, \mathbf{n}_y & R_{z}(\theta)\, \mathbf{n}_z
\end{bmatrix}.
\end{equation}
Thus, the effect of the left multiplication is to modify the representation of the body frame by rotating the lab frame about the $z$-axis by an angle $\theta$. The new rotation matrix is then given by
\begin{equation}
O' = O\,R_{z}(\theta) = N\,R_{z}(\theta).
\end{equation}
Right multiplication by $R_{z}(\theta)$ applies the rotation in the \emph{body} frame. That is, the operation rotates the body axes while keeping the global frame fixed. In particular, if the columns of $N$ are the original body axes $\mathbf{n}_x$, $\mathbf{n}_y$, and $\mathbf{n}_z$, then the updated axes (the columns of $O'$) are given by

\begin{equation}
\textbf{Intrinsic rotations:} \quad O' = \begin{bmatrix}
 \cos \theta \mathbf{n}_x -\sin \theta \mathbf{n}_y &    \cos \theta \mathbf{n}_y +\sin \theta \mathbf{n}_x     &      \mathbf{n}_z
\end{bmatrix}.
\end{equation}

Thus, the new rotation matrix $O'$ encapsulates the effect of rotating the body frame about its own $z$-axis by an angle $\theta$, while the global frame remains unchanged. Note that the rotation acts \emph{clockwise} on the axes $\mathbf{n}_x, \mathbf{n}_y$.

\section{Rotations in four dimensions}

In four-dimensional space, rotations of a vector can be classified into three main types, each with distinct characteristics:

\bigskip

\textbf{Simple Rotations:}  
A simple rotation is essentially a two-dimensional rotation embedded in 4D. Only one of the two orthogonal planes is actively rotated by a nonzero angle, while the perpendicular plane remains fixed. In other words, the rotation occurs in only one plane.

\bigskip

\textbf{Double Rotations:}  
A double rotation involves simultaneous rotations in two orthogonal planes by different angles (say, $\theta$ and $\phi$). This can be viewed as two independent rotations occurring at the same time. Unlike in 3D, there is no single axis of rotation; instead, the motion is distributed across two independent 2-planes. Double rotations are unique to four dimensions and higher. Schematically, they take the form:
\begin{equation}
\begin{pmatrix}
R(\theta) & 0 \\
0 & R(\phi)
\end{pmatrix}.
\end{equation}

\bigskip

\textbf{Isoclinic Rotations:}  
An isoclinic rotation is a special case of a double rotation where the two rotation angles are equal in magnitude (or equal up to a sign), i.e., $\theta=\phi$ or $\theta=-\phi$. This means that every vector in $\mathbb{R}^4$ is rotated through the same angle, even though there is no unique fixed axis. Schematically, an isoclinic rotation can be represented as:
\begin{equation}
\begin{pmatrix}
R(\theta) & 0 \\
0 & R(\theta)
\end{pmatrix}
\quad \text{or} \quad
\begin{pmatrix}
R(\theta) & 0 \\
0 & R(-\theta)
\end{pmatrix}.
\end{equation}
These two cases correspond to \emph{left} and \emph{right} isoclinic rotations, respectively. They are identified with the two independent actions of 
\begin{equation}
\mathfrak{so}(4) \cong \mathfrak{so}(3)_L \oplus \mathfrak{so}(3)_R.
\end{equation}

\section{$\mathfrak{so}(4)$ Algebra}

A standard way to describe the Lie algebra $\mathfrak{so}(4)$ is to introduce six generators $J_{AB}$ (with $A,B=0,1,2,3$ and $A<B$) that represent an infinitesimal rotation in the $(x_A,x_B)$-plane. Their entries are defined by
\begin{equation}
\left(J_{AB}\right)_{CD} = \delta_{AC}\delta_{BD} - \delta_{AD}\delta_{BC}.
\end{equation}
These generators are antisymmetric ($J_{AB} = -J_{BA}$) and satisfy the usual $\mathfrak{so}(4)$ commutation relations.

It is convenient to separate these generators into two groups: 
\begin{itemize}
    \item The three generators $J_{0a}$ (with $a=1,2,3$) that mix the 0-component with the ``spatial components".
    \item The three generators $J_{ab}$ (with $a,b=1,2,3$) that generate rotations in the three-dimensional ``spatial" subspace.
\end{itemize}

Explicitly, the ``boost-like" generators, which mix the 0 component with a ``spatial" index, are given by:
\begin{equation}
J_{01} = \begin{pmatrix}
0 & -1 & 0 & 0 \\
+1 & 0 & 0 & 0 \\
0 & 0 & 0 & 0 \\
0 & 0 & 0 & 0
\end{pmatrix}, \quad
J_{02} = \begin{pmatrix}
0 & 0 & -1 & 0 \\
0 & 0 & 0 & 0 \\
+1 & 0 & 0 & 0 \\
0 & 0 & 0 & 0
\end{pmatrix}, \quad
J_{03} = \begin{pmatrix}
0 & 0 & 0 & -1 \\
0 & 0 & 0 & 0 \\
0 & 0 & 0 & 0 \\
+1 & 0 & 0 & 0
\end{pmatrix}.
\end{equation}
The spatial generators, which generate rotations in the three-dimensional subspace, are:
\begin{equation}
J_{12} = \begin{pmatrix}
0 & 0 & 0 & 0 \\
0 & 0 & -1 & 0 \\
0 & +1 & 0 & 0 \\
0 & 0 & 0 & 0
\end{pmatrix}, \quad
J_{13} = \begin{pmatrix}
0 & 0 & 0 & 0 \\
0 & 0 & 0 & -1 \\
0 & 0 & 0 & 0 \\
0 & +1 & 0 & 0
\end{pmatrix}, \quad
J_{23} = \begin{pmatrix}
0 & 0 & 0 & 0 \\
0 & 0 & 0 & 0 \\
0 & 0 & 0 & -1 \\
0 & 0 & +1 & 0
\end{pmatrix}.
\end{equation}

All these generators correspond to the \textbf{simple rotations} discussed above. In particular, the $J_{0a}$ are associated with the \emph{axial} generators, $A_a$, while the $J_{1a}$ are associated with the \emph{vector} generators, $V_a$, as defined in the main text.

On the other hand, a \textbf{double rotation} is generated by a linear combination of two commuting generators that act in two orthogonal planes. For instance, if we consider the $(x_0,x_1)$ and $(x_2,x_3)$ planes, a typical generator is
\begin{equation}
G = \theta\,J_{01} + \phi\,J_{23},
\end{equation}
where $J_{23}$ is defined analogously to $J_{01}$ for the $(x_2,x_3)$-plane.

Finally, an \textbf{isoclinic rotation} is a special case of a double rotation where the two rotation angles are equal in magnitude (or equal up to a sign), i.e., $\theta=\phi$ or $\theta=-\phi$. In such rotations, every vector in $\mathbb{R}^4$ is rotated through the same angle, even though there is no unique fixed axis. Using the same two planes as before, an isoclinic rotation is generated by
\begin{equation}
G_{\pm} = \theta\left(J_{01} \pm J_{23}\right).
\end{equation}
For $G_{+}=\theta\left(J_{01}+J_{23}\right)$, exponentiation yields a rotation by $\theta$ in both planes; for $G_{-}=\theta\left(J_{01}-J_{23}\right)$, the rotation is by $\theta$ in one plane and by $-\theta$ in the other.

To formally describe the isoclinic rotations, we ``split" $\mathfrak{so}(4)$ into two parts:
\begin{equation}
\mathfrak{so}(4) \cong \mathfrak{so}(3)_L \oplus \mathfrak{so}(3)_R.
\end{equation}
This is accomplished by forming the (anti)self-dual combinations of the generators:
\begin{equation}
\lambda_a = \frac{1}{2}\Bigl(J_{0a} + \frac{1}{2}\,\epsilon_{abc}\,J_{bc}\Bigr), \quad
\rho_a = \frac{1}{2}\Bigl(J_{0a} - \frac{1}{2}\,\epsilon_{abc}\,J_{bc}\Bigr),
\end{equation}
where $\epsilon_{abc}$ is the Levi-Civita symbol. (Alternatively, one may define $\rho_a$ with an overall minus sign; the choice here is conventional so that the $\lambda_a$ and $\rho_a$ satisfy)
\begin{equation}
\left[\lambda_a, \lambda_b\right] = \epsilon_{abc}\,\lambda_c, \quad \left[\rho_a, \rho_b\right] = -\epsilon_{abc}\,\rho_c,
\end{equation}
and
\begin{equation}
\left[\lambda_a, \rho_b\right] = 0.
\end{equation}

In other words, the combination $J_{0a} + \frac{1}{2}\,\epsilon_{abc}\,J_{bc}$ generates the left-action (associated with $\lambda_a$), while the combination $J_{0a} - \frac{1}{2}\,\epsilon_{abc}\,J_{bc}$ generates the right-action (associated with $\rho_a$).

For example, for $a=1$ we have
\begin{equation}
\lambda_1 = \frac{1}{2}\Bigl(J_{01} + J_{23}\Bigr) = \frac{1}{2}\begin{pmatrix}
0 & -1 & 0 & 0 \\
+1 & 0 & 0 & 0 \\
0 & 0 & 0 & -1 \\
0 & 0 & +1 & 0
\end{pmatrix}.
\end{equation}
Similarly, one finds
\begin{equation}
\lambda_2 = \frac{1}{2}\Bigl(J_{02} + J_{31}\Bigr) = \frac{1}{2}\begin{pmatrix}
0 & 0 & -1 & 0 \\
0 & 0 & 0 & +1 \\
+1 & 0 & 0 & 0 \\
0 & -1 & 0 & 0
\end{pmatrix},
\end{equation}
and
\begin{equation}
\lambda_3 = \frac{1}{2}\Bigl(J_{03} + J_{12}\Bigr) = \frac{1}{2}\begin{pmatrix}
0 & 0 & 0 & -1 \\
0 & 0 & -1 & 0 \\
0 & +1 & 0 & 0 \\
+1 & 0 & 0 & 0
\end{pmatrix}.
\end{equation}

For the right-sector generators, we define
\begin{equation}
\begin{aligned}
\rho_1 &= \frac{1}{2}\Bigl(J_{01} - J_{23}\Bigr) = \frac{1}{2}\begin{pmatrix}
0 & -1 & 0 & 0 \\
+1 & 0 & 0 & 0 \\
0 & 0 & 0 & +1 \\
0 & 0 & -1 & 0
\end{pmatrix}, \\
\rho_2 &= \frac{1}{2}\Bigl(J_{02} - J_{31}\Bigr) = \frac{1}{2}\begin{pmatrix}
0 & 0 & -1 & 0 \\
0 & 0 & 0 & -1 \\
+1 & 0 & 0 & 0 \\
0 & +1 & 0 & 0
\end{pmatrix}, \\
\rho_3 &= \frac{1}{2}\Bigl(J_{03} - J_{12}\Bigr) = \frac{1}{2}\begin{pmatrix}
0 & 0 & 0 & -1 \\
0 & 0 & +1 & 0 \\
0 & -1 & 0 & 0 \\
+1 & 0 & 0 & 0
\end{pmatrix}.
\end{aligned}
\end{equation}

\section{Isospin spin-wave solutions}

In this subsection, we derive an exact solution for waves with zero axial charge density and nonzero vector charge density. We begin by expanding the Lagrangian density from Eq.~\eqref{spin-frame-triangular-AFM} in terms of Euler angles (following convention ``313"). The resulting Lagrangian is

\begin{equation}
\mathcal{L} = \frac{\rho}{2}\Bigl[\dot{\Theta}^2 + \sin^2\Theta\dot{\Phi}^2+ \Bigl(\dot{\Psi} + \cos\Theta\dot{\Phi}\Bigr)^2 \Bigr]
-\frac{\mu}{2}\Bigl[(\nabla \Theta)^2 + \sin^2\Theta\,(\nabla \Phi)^2 + 2\Bigl(\nabla \Psi + \cos\Theta\,\nabla \Phi\Bigr)^2\Bigr].
\end{equation}

The axial and vector charge densities are given by
\begin{equation}
\begin{aligned} 
\sigma_{A,3} = 2\rho\,\cos^2\left(\frac{\Theta}{2}\right)\left(\partial_t \Phi + \partial_t \Psi\right), \quad \sigma_{V,3} = 2\rho\,\sin^2\left(\frac{\Theta}{2}\right)\left(\partial_t \Phi - \partial_t \Psi\right).
\end{aligned}
\end{equation}
Zero axial charge density implies that $\sigma_{A,3} = 0$, which in turn requires that both $\Theta$ and $\Phi+\Psi$ remain constant. Indeed, for circularly polarized solutions with a uniform isospin (vector) charge density, the trace of the spin order parameter
\begin{equation}
\mathrm{Tr}[O] = \mathbf{e}_{a}\cdot \mathbf{n}_{a}      =4q_0^2 - 1 = 4\cos^2\left(\frac{\Theta}{2}\right)\cos\left(\frac{\Phi+\Psi}{2}\right) - 1
\end{equation}
remains constant.

To obtain plane-wave solutions that reduce to the circular polarization $\phi_x \pm i\,\phi_y$ in the small-amplitude limit, we adopt the following ansatz:
\begin{equation}
\begin{aligned}
\Theta(t,\mathbf{r}) &= \Theta_0, \\
\Phi(t,\mathbf{r}) &= \omega t - \mathbf{k}\cdot\mathbf{r} + \Phi_0, \\
\Psi(t,\mathbf{r}) &= -\omega t + \mathbf{k}\cdot\mathbf{r} + \Psi_0,
\end{aligned}
\end{equation}
where $\Theta_0$, $\Phi_0$, and $\Psi_0$ are constants.

\begin{equation}
\begin{aligned}
\rho\Bigl( \ddot{\Theta} + \sin\Theta\dot{\Phi}\dot{\Psi} \Bigr)
- \mu\nabla^2\Theta + \mu\sin\Theta\cos\Theta(\nabla\Phi)^2 - 2\mu\sin\Theta\Bigl[(\nabla\Psi+\cos\Theta\nabla\Phi)\cdot\nabla\Phi\Bigr] &= 0\,, \\[1mm]
\rho\partial_{t}\Bigl[ \partial_{t}\Phi + \cos\Theta\partial_{t}\Psi \Bigr]
- \mu\nabla\cdot\Bigl[\sin^2\Theta\nabla\Phi + 2\cos\Theta\Bigl(\nabla\Psi+\cos\Theta\nabla\Phi\Bigr)\Bigr] &= 0, \\[1mm]
\rho\partial_{t}\Bigl[ \partial_{t}\Psi + \cos\Theta\partial_{t}\Phi \Bigr]
- 2\mu\nabla\cdot\Bigl(\nabla\Psi+\cos\Theta\nabla\Phi\Bigr) &= 0.
\end{aligned}
\end{equation}

Since $\Phi$ and $\Psi$ are linear in time, their equations of motion are automatically satisfied; they reduce to the conservation of currents. Notice that
\begin{equation}
\partial_t\Phi + \cos\Theta_0\,\partial_t\Psi =  (1-\cos\Theta_0)\omega , \quad   \sin^2\Theta\,\nabla\Phi + 2\cos\Theta\Bigl(\nabla\Psi+\cos\Theta\,\nabla\Phi\Bigr)
= -(1-\cos\Theta_0)^2\mathbf{k}.
\end{equation}
and 
\begin{equation}
   \partial_t\Psi + \cos\Theta_0\,\partial_t\Phi =  (\cos\Theta_0 - 1)\omega, \quad  \nabla\Psi + \cos\Theta_0\,\nabla\Phi  = (1-\cos\Theta_0 )\mathbf{k}.
\end{equation}

The only no trivial equation corresponds to the equation of motion for $\Theta$, which reduces to
\begin{equation}
\rho\,\omega^2 \sin\Theta_0 - \mu\,|\mathbf{k}|^2 \sin\Theta_0\Bigl(2-\cos\Theta_0\Bigr) = 0.
\end{equation}
For nontrivial solutions (with $\sin\Theta_0 \neq 0$), we obtain the dispersion relation:
\begin{equation} 
\omega^2 = \Bigl(2-\cos\Theta_0\Bigr)c_{\mathrm{I}}^2 |\mathbf{k}|^2,
\end{equation}
where $c_{\mathrm{I}}^2 = \mu/\rho$. The effective ``stiffness" is modified by the wave amplitude $\Theta_{0}$. In the limit of vanishing amplitude, $\Theta_0\rightarrow 0$, we recover
\begin{equation}
\omega^2 = c_{\mathrm{I}}^2 |\mathbf{k}|^2.
\end{equation}

In the language of rotation matrices, the order parameter for these solutions is constructed as the conjugation of a rotation about the $x$-axis, $R_x(\Theta_{0})$, by a rotation about the $z$-axis:
\begin{equation}
    O = R_z(\Phi)\, R_x(\Theta_{0})\, R_z^{-1}(\Phi).
\end{equation}
This operation rotates the $x$-axis into a new direction. Specifically, applying $R_z(\Phi)$ to the unit vector $(1,0,0)^T$ yields
\begin{equation} 
    \hat{\mathbf{n}} = R_z(\Phi)(1,0,0)^T = (\cos \Phi, \sin \Phi, 0)^T, 
\end{equation}
so that the net effect is equivalent to a rotation by an angle $\Theta_{0}$ about the axis $\mathbf{n} = (\cos \Phi, \sin \Phi, 0)^T$ within the spin plane defined by the ground state. Since $\Phi = \omega t - \mathbf{k}\cdot\mathbf{x}$, this axis rotates in the spin plane with a constant angular velocity $\omega$.

A useful identity connecting $\mathrm{SO(3)}$ matrices and $\mathrm{SU(2)}$ is given by
\begin{equation} 
    O_{ab} = \Bigl(2q_0^2 - 1\Bigr) \delta_{ab} + 2q_aq_b - 2q_0\epsilon_{abc}q_c,
\end{equation}
where $a,b,c = 1,2,3$, $\delta_{ab}$ is the Kronecker delta, and $\epsilon_{abc}$ is the Levi-Civita symbol. More explicitly, the rotation matrix can be written as
\begin{equation}
O(\mathbf{q}) =
\begin{bmatrix}
2\left(q_0^2+q_1^2\right)-1 & 2\left(q_1q_2-q_0q_3\right) & 2\left(q_1q_3+q_0q_2\right) \\
2\left(q_1q_2+q_0q_3\right) & 2\left(q_0^2+q_2^2\right)-1 & 2\left(q_2q_3-q_0q_1\right) \\
2\left(q_1q_3-q_0q_2\right) & 2\left(q_2q_3+q_0q_1\right) & 2\left(q_0^2+q_3^2\right)-1
\end{bmatrix}.
\end{equation}

Using the above identity, the trace of the spin order parameter $R$ is given by
\begin{equation}
    \mathrm{Tr}[O] = 4q_0^2 - 1 = 4\cos^2\left(\frac{\Theta}{2}\right)\cos\left(\frac{\Phi+\Psi}{2}\right) - 1,
\end{equation}
which remains constant for solutions with well-defined isospin density. Assuming that $\Phi_0+\Psi_0=0$ (or, equivalently, absorbing any constant phase into the redefinition of $\Theta_{0}$), we have 
\begin{equation}
\begin{aligned}
    \mathrm{Tr}[O] &= \mathbf{e}_{a}\cdot\mathbf{n}_{a} = 1+2\cos \Theta_{0} \\
    &\approx 3-\Theta_{0}^2,
\end{aligned}
\end{equation}
for small-amplitude spin waves. Therefore, near the ground state, the trace of $R$ provides a direct measure of the spin-wave amplitude.

\section{$ \mathrm{SO(3)_{L}}\times \mathrm{SO(2)_{R,3}}$-Invariance}

The tangent vectors $e_a$, which span the local basis in four-dimensional space, are defined by
\begin{equation}
    e_a = 2 \rho_a q.
\end{equation}
Under an infinitesimal extrinsic rotation with parameters $\phi_b$, the order parameter $q$ transforms as
\begin{equation}
    q \mapsto q + \phi_b \lambda_b q.
\end{equation}
Because the intrinsic generators $\rho_a$ commute with the global generators $\lambda_b$, the tangent vectors transform as
\begin{equation}
\begin{aligned}
    e_a &\mapsto 2 \rho_a q' = 2 \rho_a \bigl(q + \phi_b \lambda_b q\bigr)\\
        &= e_a + \phi_b \lambda_b e_a.
\end{aligned}
\end{equation}
Thus the local basis vectors $e_a$ transform in the same way as the order parameter $q$.

By contrast, under intrinsic rotations (generated by $\rho_a$), the order parameter transforms as a generic vector in four dimensions:
\begin{equation}
    q \mapsto q + \psi_b\,\rho_b\,q.
\end{equation}
The tangent vectors $e_a = 2\,\rho_a\,q$ then transform as
\begin{equation}
\begin{aligned}
    e_a &\mapsto 2\rho_a q' 
         = 2\rho_a\bigl(q + \psi_b \rho_b q\bigr)\\
        &= e_a + 2\psi_b \rho_a \rho_b q
         = e_a + \psi_b \rho_a e_b.
\end{aligned}
\end{equation}

The $\mathrm{O(4)}$-NLSM that describes the triangular antiferromagnet is given by
\begin{align} \label{O(4)-NLSM-triangular}
\mathcal{L} ={}& 2\rho \left[ (e_1 \cdot \partial_t q)^2 + (e_2 \cdot \partial_t q)^2 + (e_3 \cdot \partial_t q)^2 \right] \nonumber \\
& - 2\mu \left[ (e_1 \cdot \nabla q)^2 + (e_2 \cdot \nabla q)^2 + 2(e_3 \cdot \nabla q)^2 \right],
\end{align}
where $q \in S^3$ is the four-component spinor and $\{e_a\}$ are local orthonormal vectors orthogonal to $q$. We now investigate how this Lagrangian transforms under extrinsic and intrinsic $\mathrm{SO(4)}$ rotations.

The kinetic term in Eq.~\eqref{O(4)-NLSM-triangular} is manifestly invariant under both extrinsic and intrinsic rotations. The spatial-gradient term, however, behaves differently. Under an infinitesimal transformation, the variation of each spatial term is given by
\begin{equation}
\delta (e_a \cdot \nabla q)^2 = 2 (e_a \cdot \nabla q) \, \delta (e_a \cdot \nabla q),
\end{equation}
so we must compute $\delta(e_a \cdot \nabla q)$. Under an infinitesimal extrinsic rotation generated by $\lambda_b$, we find:
\begin{align}
\delta_{\lambda_b} (e_a \cdot \nabla q)
&= (\delta_{\lambda_b} e_a) \cdot \nabla q + e_a \cdot (\delta_{\lambda_b} \nabla q) \nonumber \\
&= \phi_b \lambda_b e_a \cdot \nabla q + e_a \cdot (\phi_b \lambda_b \nabla q) \nonumber \\
&= \phi_b \left( \lambda_b e_a \cdot \nabla q - \lambda_b e_a \cdot \nabla q \right) = 0,
\end{align}
where we have used the antisymmetry of the generators: $\lambda_b^{-1} = -\lambda_b$. It follows that the Lagrangian~\eqref{O(4)-NLSM-triangular} is invariant under global (extrinsic) $\mathrm{SO(3)_L}$ rotations:
\begin{equation}
\delta_{\lambda_b} \mathcal{L} = 0.
\end{equation}

Next, we consider intrinsic rotations generated by $\rho_b$. In this case:
\begin{align}
\delta_{\rho_b} (e_a \cdot \nabla q)
&= (\delta_{\rho_b} e_a) \cdot \nabla q + e_a \cdot (\delta_{\rho_b} \nabla q) \nonumber \\
&= 2\psi_b (\rho_a \rho_b q) \cdot \nabla q - 2\psi_b (\rho_b \rho_a q) \cdot \nabla q \nonumber \\
&= 2\psi_b \left[ (\rho_a \rho_b - \rho_b \rho_a) q \right] \cdot \nabla q \nonumber \\
&= -2\psi_b \epsilon_{abc} (\rho_c q) \cdot \nabla q \nonumber \\
&= -\epsilon_{abc} \psi_b \, e_c \cdot \nabla q.
\end{align}

where we used the identity $\rho_a q \cdot (\rho_b \nabla q) = \rho_b^{-1} \rho_a q \cdot \nabla q$ and that $\rho_b^{-1} = -\rho_b$. Rewriting Eq.~\eqref{O(4)-NLSM-triangular} as
\begin{equation}
\mathcal{L} = \mathcal{L}_{\mathrm{PCM}} - 2\mu (e_3 \cdot \nabla q)^2,
\end{equation}
where $\mathcal{L}_{\mathrm{PCM}}$ is the fully $\mathrm{SO(4)}$-symmetric principal chiral model. The variation of the full Lagrangian under intrinsic rotations becomes
\begin{equation}
\delta_{\rho_b} \mathcal{L} = \underbrace{\delta_{\rho_b} \mathcal{L}_{\mathrm{PCM}}}_{=0}
- 4\mu (e_3 \cdot \nabla q) \, \delta_{\rho_b}(e_3 \cdot \nabla q).
\end{equation}
This last term vanishes only if $b = 3$, that is, when the intrinsic rotation is about the third axis, $\mathbf{n}_{z}$. Thus, the symmetry is reduced from $\mathrm{SO(3)_R}$ to $\mathrm{SO(2)_{R,3}}$, preserving only rotations about the spin-plane normal.

\end{widetext}

\end{document}